\documentclass[pra,aps,twocolumn,nofootinbib,noshowpacs,preprintnumbers,longbibliography,floatfix,groupedaddress]{revtex4-2}

\usepackage[utf8]{inputenc}
\usepackage{graphicx}
\usepackage{float}
\usepackage{amssymb}
\usepackage{amsmath}  
\usepackage{mathtools}
\usepackage{dsfont}
\usepackage{array}
\usepackage{bm,fixmath}
\usepackage{mathrsfs}
\usepackage{pifont}
\usepackage{multirow}
\usepackage{upgreek}
\usepackage{xcolor}
\usepackage{bm}
\usepackage{bbm}
\usepackage{physics}
\usepackage{comment}
\usepackage{appendix}
\usepackage{verbatim}
\usepackage{slashed}
\usepackage[pdftex,
            colorlinks,
            linkcolor=red,
            citecolor=blue,
            urlcolor=black]{hyperref}
\usepackage{cleveref}
\usepackage{comment}
\usepackage[caption=false]{subfig}

\newcommand{\bellstate}[2]{\ket{\psi_{#1 #2}}}
\newcommand{\mbellstate}[3]{\ket{\Psi^{#3}_{#1 #2}}}

\newcommand{\no}[1]{\text{\sout{\ensuremath{#1}}}}

\definecolor{cyan}{rgb}{0.0, 1.0, 1.0}

\usepackage{tikz,esvect,tikz-3dplot}
\usetikzlibrary{3d,calc,intersections}
\usepackage{tikz}
\usepackage{tikz-3dplot}
\usepackage{quantikz}
\usepackage{lipsum}
\usepackage{tkz-graph}
\usetikzlibrary{positioning}
\pgfdeclarelayer{background}
\pgfdeclarelayer{foreground}
\pgfsetlayers{background,main,foreground}
\newcommand{\mathcolorbox}[2]{\colorbox{#1}{$\displaystyle #2$}}

\usepackage[normalem]{ulem}

\begin{document}

\title{High photon-loss threshold quantum computing using GHZ-state measurements}

\author{Brendan Pankovich}
\author{Angus Kan}
\author{Kwok Ho Wan}
\author{Maike Ostmann}
\author{Alex Neville}
\author{Srikrishna Omkar}
\author{Adel Sohbi}
\author{Kamil Brádler}
\email{kamil@orcacomputing.com}
\affiliation{ORCA Computing}

\begin{abstract}
We propose fault-tolerant architectures based on performing projective measurements in the Greenberger–Horne–Zeilinger (GHZ) basis on constant-sized, entangled resource states. We present linear-optical constructions of the architectures, where the GHZ-state measurements are encoded to suppress the errors induced by photon loss and the probabilistic nature of linear optics. Simulations of our constructions demonstrate high single-photon loss thresholds compared to the state-of-the-art linear-optical architecture realized with encoded two-qubit fusion measurements performed on constant-sized resource states. We believe this result shows a resource-efficient path to achieving photonic fault-tolerant quantum computing.
\end{abstract}

\maketitle

\section{Introduction}

Fault-tolerant quantum computation relies on effective correction of hardware errors during the execution of a quantum computer program. Practical implementations of fault tolerance depend largely on the features of the underlying hardware. For instance, circuit-based error correction provides a framework for hardware equipped with deterministic gates to detect errors by performing non-destructive, ancilla-assisted measurements~\cite{fowler2012surface,terhal_qec}. In measurement-based quantum computing (MBQC), on the other hand, error syndromes are constructed from destructive measurements performed on previously generated entangled states. MBQC is well-established to be suitable for hardware with probabilistic entangling operations and destructive measurements~\cite{raussendorf2007topological,barrett2010fault,auger2018fault,bartolucci2023fusion}, such as discrete variable photonic qubits~\cite{gimeno2015from,li2015resource,omkar2022all,bartolucci2023fusion} and continuous variable qubits~\cite{fukui2018high,Bourassa2021blueprintscalable,larsen2021fault}.

Many photonic MBQC architectures~\cite{fujii2010fault,li2010fault,herrera2010photonic,barrett2010fault,gimeno2015from,li2015resource,auger2018fault,pant2019percolation,omkar2022all} achieve fault tolerance in two stages: (i) preparation of a large entangled resource state, whose size grows with the desired quantum computer program, followed by (ii) destructive single-qubit measurements on the prepared state to execute the program. A streamlined approach to MBQC, recently proposed in Ref.~\cite{bartolucci2023fusion}, called fusion-based quantum computation (FBQC) performs destructive, two-qubit projective measurements in the Bell-state basis, known as Bell-state measurements~\cite{linear2007kok} (BSMs) or fusions~\cite{resource2005browne}, on constant-sized resource states. Further reported in Ref.~\cite{bartolucci2023fusion} are FBQC architectures that implement the surface code with high thresholds for photon loss and fusion failures. Similar thresholds have since been obtained for other topological error correction codes implemented using FBQC~\cite{sahay2023high,paesani2022high}. The high thresholds and rapidly progressing experimental capabilities to generate entangled photonic resource states~\cite{schwartz2016deterministic,wang2016experimental,istrati2020sequential,thomas2022efficient,maring2023general} suggest that photonic platforms are suitable for implementing FBQC architectures. Yet, the size of the resource states and the photon loss thresholds demanded by these architectures still remain challenging to be reached by current hardware.

In this paper, we explore alternative photonic architectures with the aim of alleviating these demands on the hardware. Specifically, we devise MBQC architectures that achieve fault tolerance by fusing $n$ resource states using measurements in the $n$-qubit Greenberger–Horne–Zeilinger (GHZ) state basis, or GHZ-state measurements (GSMs). In order to provide practical photonic realizations of our architectures, we construct encoded GSMs, where the encoding suppresses measurement failures and the effects of photon loss, from a collection of physical BSMs. We apply the architectures to implement the surface code, and numerically obtain high photon-loss thresholds. In particular, we present two families of constant-sized, encoded two-qubit graph states with high single-photon loss thresholds. We then compare our results to the 24-qubit, encoded six-ring FBQC construction~\cite{bartolucci2023fusion}. We find that (1) resource states with the same number of photons can achieve a $\sim 75\%$ improvement in the single-photon loss threshold, and (2) resource states that are $ 1/3$ smaller can achieve a similar loss threshold. Note that in this work, we focus on how to use such resource states to perform fault-tolerant quantum computation. Photonic resource state generation methods can be found in Refs.~\cite{physicalghz,resource2005browne,ewert2017ultrafast,istrati2020sequential,bartolucci2021creation,omkar2022all}.

The rest of the paper is organized as follows. We present our architectures in section~\ref{sec:construction}, wherein we define the entangling measurements, resource states, and error correction methods. In section~\ref{sec:performance}, we present linear-optical implementations of the encoded GSMs, and the simulated photon-loss thresholds. Finally, we discuss our findings and provide an outlook in section~\ref{sec:discussion}.

\section{GHZ measurement-based architectures}
\label{sec:construction}

In fault-tolerant MBQC, error syndrome data are constructed from projective measurements performed on a set of entangled, resource states. Formally, the resource states, specified by their stabilizer group~\cite{gottesman1997stabilizer} $\mathcal{R}$, are projected onto the basis of a commuting set of Pauli observables, which generate another stabilizer group $\mathcal{M}$. Then, the error syndromes come from the check operator group $\mathcal{C} = \mathcal{R}\cap\mathcal{M}$. The generators of $\mathcal{C}$ known as parity check operators are used to detect errors~\cite{raussendorf2007topological,fowler2009topological,brown2020universal,bartolucci2023fusion}. 

A process called \emph{foliation}~\cite{bolt2016foliated,brown2020universal} is commonly used to configure the resource states and measurements. Specifically, foliation generates a graph from a stabilizer error correction code. The graph can be used directly as the defining entanglement structure of the target graph state~\cite{raussendorf2001a}, on which single-qubit Pauli measurements are performed for fault tolerance~\cite{bolt2016foliated,brown2020universal}. 

Alternatively, the graph can be used as a coordinate system known as a \emph{fusion network}~\cite{bartolucci2023fusion} that determines the locations of resource states and entangling measurements. In our architectures, we consider the Raussendorf-Harrington-Goyal (RHG) lattice~\cite{raussendorf2007topological}, or equivalently, the foliated surface code~\cite{bolt2016foliated} as a fusion network. We use two-qubit linear-chain graph states, of which $X_1 Z_2$ and $Z_1 X_2$ are stabilizers, where $X_i$ and $Z_i$ are Pauli-x and -z operators acting on the $i$th qubit respectively, as our resource states. We place them along the edges of a RHG lattice, and end up with $n$ qubits at each vertex, where $n$ is the degree of the vertex.
Next, we perform an $n$-qubit GSM at each vertex. In a RHG lattice, vertices have degree four in the bulk and less than four at the boundary~\cite{fowler2009topological}. As such, our architecture is dual to the 4-star FBQC construction~\cite{bartolucci2023fusion}, where in the bulk, 4-qubit GHZ states placed at the vertices are jointly measured along the edges in the basis of $X_1 Z_2$ and $Z_1 X_2$.

\subsection{Entangling measurements}
\label{sec:encoded_GHZ}

\begin{figure}[!t]
\center
\resizebox{\columnwidth}{!}{%
\begin{tikzpicture}[scale=.45,every node/.style={minimum size=0.2cm},on grid]

    \fill[white,fill opacity=1] (0,0) rectangle (10,6);

    \fill[white!90!black] (4,8) ellipse ({2} and {1}); 
    \draw (4,8) ellipse [x radius=2,y radius=1];
    \node[fill=white,opacity=1,shape=circle,draw=black] (n2b) at (3,8) {$2b$};
    \node[fill=white,opacity=1,shape=circle,draw=black] (n2a) at (5,8) {$2a$};

    \fill[white!90!black] (0,4) ellipse ({1} and {2}); 
    \draw (0,4) ellipse [x radius=1,y radius=2];
    \node[fill=white,opacity=1,shape=circle,draw=black] (n3b) at (0,3) {$3b$};
    \node[fill=white,opacity=1,shape=circle,draw=black] (n3a) at (0,5) {$3a$};

    \path[shape=coordinate]
        (7,6) coordinate(b1) (9,6) coordinate(b2)
        (9,2) coordinate(b3) (7,2) coordinate(b4);
    \filldraw[fill=white!90!black] (b1) -- (b2) -- (b3) -- (b4) -- (b1);
    \node[fill=white,opacity=1,shape=circle,draw=black] (n1b) at (8,5) {$1b$};
    \node[fill=white,opacity=1,shape=circle,draw=black] (n1a) at (8,3) {$1a$};

    \path[shape=coordinate]
        (2,1) coordinate(c1) (6,1) coordinate(c2)
        (6,-1) coordinate(c3) (2,-1) coordinate(c4);
    \filldraw[fill=white!90!black] (c1) -- (c2) -- (c3) -- (c4) -- (c1);
    \node[fill=white,opacity=1,shape=circle,draw=black] (n4a) at (3,0) {$4a$};
    \node[fill=white,opacity=1,shape=circle,draw=black] (n4b) at (5,0) {$4b$};

    \draw [orange,fill=orange,opacity=0.3] (n3a.north) -- plot [smooth] coordinates {(n2b.west) (n2b.north) (n2b.east) (n2b.south) (n3a.east) (n3a.south) (n3a.west) (n3a.north)} --cycle;
    \draw [orange,fill=orange,opacity=0.3] (n3b.south) -- plot [smooth] coordinates {(n4a.west) (n4a.south) (n4a.east) (n4a.north) (n3b.east) (n3b.north) (n3b.west) (n3b.south)} --cycle;

    \draw [orange,fill=orange,opacity=0.3] (n1b.north) -- plot [smooth] coordinates {(n2a.east) (n2a.north) (n2a.west) (n2a.south) (n1b.west) (n1b.south) (n1b.east) (n1b.north)} --cycle;

    \draw [cyan,fill=cyan,opacity=0.3] (n4b.east) -- plot [smooth] coordinates {(n1a.south) (n1a.east) (n1a.north) (n1a.west) (n4b.north) (n4b.west) (n4b.south) (n4b.east)} --cycle;

    \node[] (BSM) at (10.5,9) {\text{BSM:}};
    \node[] (BSM_stab1) at (16,8) {$\Big\langle \mathcolorbox{orange!30}{X_{1b}X_{2a}}, \mathcolorbox{orange!30}{X_{2b}X_{3a}},\mathcolorbox{orange!30}{X_{3b}X_{4a}},\mathcolorbox{cyan!30}{X_{4b}X_{1a}},$};
    
    \node[] (BSM_stab2) at (16.5,7) {$\mathcolorbox{orange!30}{Z_{1b}Z_{2a}},\mathcolorbox{orange!30}{Z_{2b}Z_{3a}},\mathcolorbox{orange!30}{Z_{3b}Z_{4a}}, \mathcolorbox{cyan!30}{Z_{4b}Z_{1a}}\Big\rangle$};

    \node[scale=3] (Darrow) at (16.5,4.5) {$\Downarrow$};

    \node[] (GSM) at (10.5,2) {\text{GSM:}};
    \node[] (GSM_stab) at (16.5,1) {$\Big\langle \mathcolorbox{orange!30}{X_{1}X_{2}X_{3}X_{4}}, \mathcolorbox{orange!30}{Z_{1}Z_{2}},\mathcolorbox{orange!30}{Z_{2}Z_{3}},\mathcolorbox{orange!30}{Z_{3}Z_{4}},\mathcolorbox{cyan!30}{Z_{4}Z_{1}}\Big\rangle$}; 
\end{tikzpicture}}
\caption{\textbf{A 4-qubit GHZ-state measurement (GSM).} \emph{Minimal} 4-qubit GSM: Qubits 2 and 3 are encoded in a two-qubit repetition code, whereas qubit 1 and 4 are not, i.e., the qubit pair highlighted in cyan do not exist, and the labels $1b$ and $4a$ are taken to be $1$ and $4$. Each qubit pair highlighted in orange is combined at a Bell-state measurement (BSM). The GSM outcomes highlighted in orange are derived from the BSM outcomes highlighted in orange. \emph{Cyclic} 4-qubit GSM: All qubits are encoded in a two-qubit repetition code. An additional BSM, highlighted in cyan, is included. As a result, two extra BSM outcomes and one extra GSM outcome, highlighted in cyan, are produced. We use ellipses to depict the qubits that are encoded in both the minimal and cyclic GSM, and rectangles to depict those that are encoded only in the cyclic GSM.}
\label{fig:4GHZ}
\end{figure}
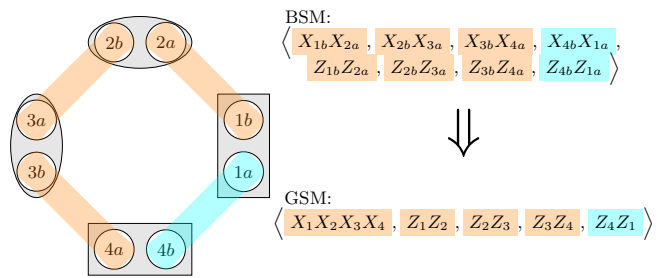

Here we define the entangling measurements, namely GSMs, in our architectures. In our first architecture, each $n$-qubit GSM measures the Pauli operators in the GHZ basis $\{ \prod_{i=1}^n X_i , Z_{j}Z_{j+1} \}_{j=1}^{n-1}$. In our second architecture, we measure an additional operator $Z_{n}Z_1$ per GSM. Henceforth, we refer to the first and second type of GSM as \emph{minimal} and \emph{cyclic} GSM, respectively, and their associated architectures as \emph{minimal} and \emph{cyclic} architectures. Furthermore, we construct both types of GSMs using at most $n$ BSMs, which measure two-qubit Pauli operators $X_1 X_2$ and $Z_1 Z_2$. For example, we provide a depiction of both types of 4-qubit GSM in Fig.~\ref{fig:4GHZ}.
While this construction is motivated by the fact that BSM is a natural entangling operation in linear optics, it is applicable to any universal hardware.

In the $n$-qubit \emph{minimal} GSM, we first encode the $i$th qubit, for $1<i<n$, in a two-qubit repetition code, which has the stabilizer $Z_{ia} Z_{ib}$ where $a$ and $b$ label the physical qubits. As such, the logical operators of the encoded qubits are given by
\begin{equation}\label{eq:2qrep}
    Z_i = Z_{ia} \sim Z_{ib}, \: X_i = X_{ia}X_{ib},
\end{equation}
where the symbol $(\sim)$ represents logical equivalence. Then, we perform $n-1$ BSMs on qubits $ib$ and $(i+1)a$ for $1<i<n-1$, as well as qubits $1$ and $2a$, and qubits $(n-1)b$ and $n$. In other words, we measure the operators in $\mathcal{O}_m = \{ X_{ib}X_{(i+1)a}, Z_{ib}Z_{(i+1)a}\}_{i=2}^{n-2} \cup \{ X_{1}X_{2a}, Z_{1}Z_{2a}, X_{(n-1)b}X_{n}, Z_{(n-1)b}Z_{n}\}$. Using the measurement outcomes and Eqn.~\eqref{eq:2qrep}, we can deduce the eigenvalues of the encoded GHZ basis. Specifically, each $ZZ$ operator in $\mathcal{O}_m$ is equivalent to a logical operator $Z_i Z_{i+1}$, and that the product of all $XX$ operators in $\mathcal{O}_m$ is equivalent to the logical operator $\prod_{i=1}^n X_i$. In the minimal architecture, the measurement stabilizer group $\mathcal{M}_m$ is a $q$-fold tensor product of $\mathcal{O}_m$ where $q$ is the number of GSMs in the fusion network.

In the $n$-qubit \emph{cyclic} GSM, we additionally encode qubits 1 and $n$ in a two-qubit repetition code. Moreover, we perform $n$ BSMs on qubits $ib$ and $(i+1)a$ for $1\leq i \leq n$, where addition is performed modulo $n$. The measurement outcomes are the eigenvalues of the operators in $\mathcal{O}_c = \{ X_{ib}X_{(i+1)a}, Z_{ib}Z_{(i+1)a}\}_{i=1}^{n}$. Once again using Eqn.~\eqref{eq:2qrep}, one can show that the $Z_{ib}Z_{(i+1)a}$ operators in $\mathcal{O}_c$ are equivalent to the logical operators $Z_i Z_{i+1}$, and the product of all $XX$ operators in $\mathcal{O}_c$ is equivalent to the logical operator $\prod_{i=1}^n X_i$. Similar to the minimal architecture, $\mathcal{O}_c$ is the generating set of $\mathcal{M}_c$ in the cyclic architecture. Since $\mathcal{O}_c$ is larger than $\mathcal{O}_m$, for a given fusion network, the cyclic architecture has a larger measurement stabilizer group than the minimal architecture.

\subsection{Resource states}
\label{sec:resource}

\begin{figure}[]
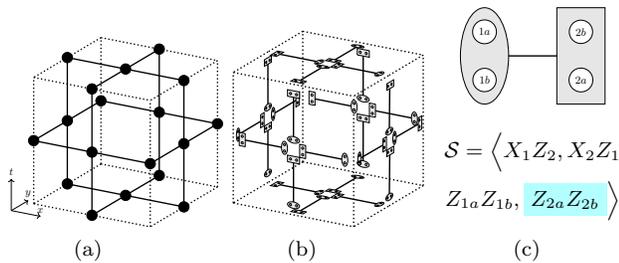

\centering
\hspace{-0.9cm}
\subfloat[]{
	\label{subfig:cube}
	\newcommand{\viewa}{30}
    \newcommand{\viewb}{20}
    \tdplotsetmaincoords{90-\viewb}{90+\viewa}
\resizebox{0.34\columnwidth}{!}{
}}
\caption{(a) A unit cell of a RHG lattice. Here, $x,y$ are the spatial dimensions and $t$ is the time dimension. (b) A resource state lay-out on a unit cell. The resource states are placed along the edges of a unit cell of a RHG lattice. (c) Each resource state is a two-qubit graph state, where qubit 1 is encoded in a two-repetition code and qubit 2 is encoded in the cyclic architecture, but not in the minimal architecture, i.e., qubit $2b$ does not exist and the label $2a$ is taken to be $2$. This is reflected in its stabilizers $\mathcal{S}$, in which the operator highlighted in cyan is in $\mathcal{S}$ only in the cyclic architecture, and not the minimal architecture. In accordance with Fig.~\ref{fig:4GHZ}, we use different shapes to represent the encoded and unencoded qubits in a minimal GSM.}
\label{fig:unit_cell}
\end{figure}

In order to render the resource states compatible with the encoded GSMs described above, we also need to encode the resource states. In the minimal architecture, we encode one qubit, out of each two-qubit graph state, in a two-qubit repetition code. Furthermore, we choose a resource-state layout that ensures in each minimal GSM, only two partaking qubits are not encoded, whereas the rest are encoded. We provide an example layout in Fig.~\ref{fig:unit_cell} and further details in App.~\ref{app:sim}. As such, the stabilizers of a resource state can be written as $\mathcal{S}_m = \{ X_{1a}X_{1b} Z_2, Z_{1a} X_2, Z_{1a}Z_{1b}\}$, where we have used Eqn.~\eqref{eq:2qrep} and assumed WLOG qubit 1 is encoded. In the cyclic architecture, both qubits in each two-qubit graph state must be encoded, since cyclic GSMs admit only encoded qubits. Then, the stabilizers of a resource state can be expressed as $\mathcal{S}_c = \{ X_{1a}X_{1b} Z_{2a}, Z_{1a} X_{2a}X_{2b}, Z_{1a}Z_{1b}, Z_{2a}Z_{2b}\}$. We remark that the resource-state stabilizer group $\mathcal{R}$ of the minimal and cyclic architectures are $k$-fold tensor products of $\mathcal{S}_m$ and $\mathcal{S}_c$, respectively, where $k$ is the number of resource states. Therefore, for a given fusion network, the cyclic architecture has not only a larger $\mathcal{M}$, but also a larger $\mathcal{R}$. The consequent difference in the check group $\mathcal{C}$ will be discussed below. 

\subsection{Error correction}
\label{sec:QEC}

In measurement-based implementations of topological error correction codes such as the surface code, a measurement outcome supports two parity check operators\footnote{If the code has boundaries and the outcome is at a boundary, then the outcome supports one parity check operator~\cite{fowler2009topological} See App.~\ref{app:sim} for details.}~\cite{raussendorf2007topological,fowler2009topological,bartolucci2023fusion}, which are stabilizers that generate the check operator group $\mathcal{C}$. This enables one to perform error correction for both erasure and Pauli errors in the measurement outcomes. We will illustrate the error correction methods using the syndrome graph representation, where measurement outcomes and parity check operators are represented as edges and vertices, respectively. The value of an edge is given by its corresponding measurement outcome $\in \{ \pm 1 \}$; the value of a vertex is the product of the values of all incident edges. If a measurement outcome is erased, we no longer have enough information to deduce the sign of its supporting parity check operators. In a syndrome graph, an erasure error is handled by removing the erased edge and then, merging the vertices at its endpoints into one vertex, which represents a stabilizer in $\mathcal{C}$ that is independent of the erased measurement outcome~\cite{barrett2010fault}. A Pauli error flips the sign of an edge and the vertices connected by the edge. It follows that a chain of Pauli errors flip the two vertices at its endpoints. This property makes a minimum weight perfect matching algorithm~\cite{raussendorf2007topological,pymatchingv2} a suitable decoder. In practice, the syndrome graph is first modified to reflect all the erasure errors. Then, Pauli errors are applied to the modified graph before processing it with a decoder.

\begin{figure*}[!htbp]
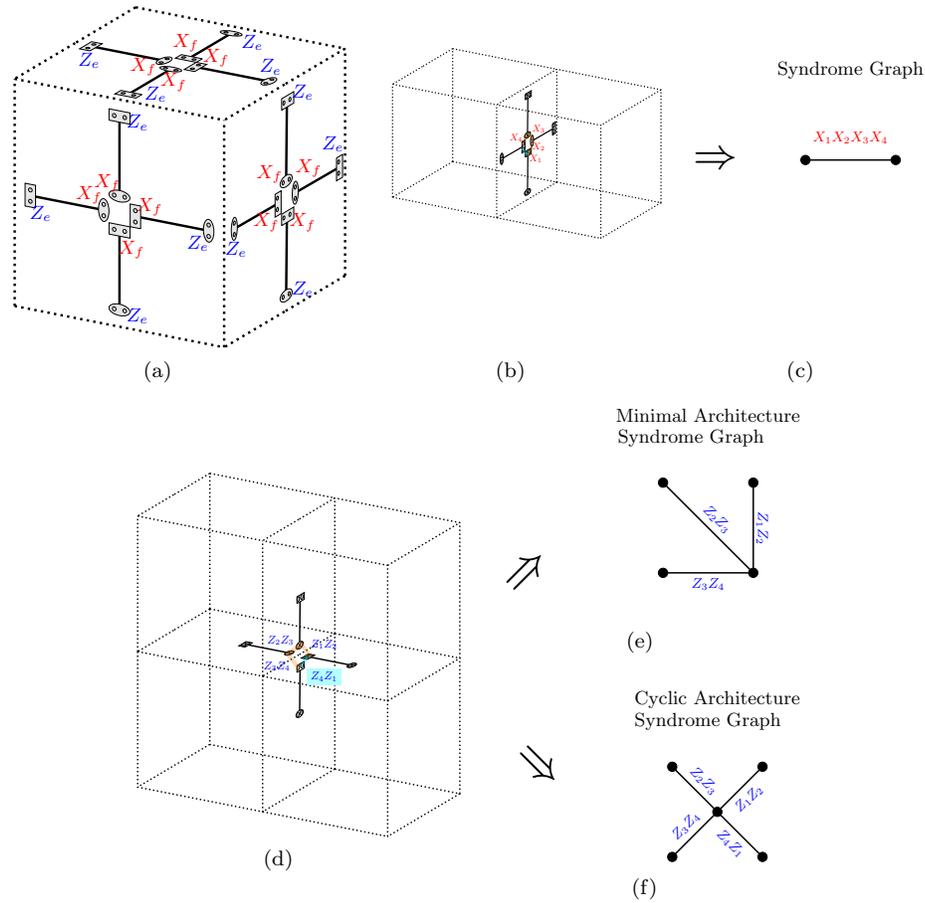

\centering
\subfloat[]{    \centering
    
    \newcommand{\viewa}{30}
    \newcommand{\viewb}{20}

    \tdplotsetmaincoords{90-\viewb}{90+\viewa}
    \resizebox{0.25\linewidth}{!}{
}
\end{minipage}

\caption{(a) The parity check operator defined on a unit cell. It is a product of $X$ operators on the face qubits and $Z$ operators on the edge qubits. (b) A GHZ-state measurement at a face of a unit cell produces a measurement outcome $X_1 X_2 X_3 X_4$, which supports two parity check operators sharing the face. We only show the resource states that are measured here. (c) The syndrome graph representation of (b), where the horizontal edge and the two vertices represent the $X_1 X_2 X_3 X_4$ outcome, and its supporting parity check operators, respectively. (d) In the \emph{minimal} architecture, a GHZ-state measurement at an edge of a unit cell produces three $ZZ$-type measurement outcomes, $Z_1 Z_2, Z_2 Z_3$, and $Z_3 Z_4$, which support the top-right, top-left, and bottom-left parity check operators, respectively. The three outcomes also support the bottom-right parity check operator because their product, i.e., $Z_4 Z_1 = Z_1 Z_2 Z_2 Z_3 Z_3 Z_4$ supports it. In the \emph{cyclic} architecture, a GHZ-state measurement at an edge directly produces the outcome $Z_4 Z_1$, which supports the bottom-right parity check operator. Moreover, the product of all four $ZZ$-type outcomes form a different parity check operator that is independent from the ones defined on unit cells. Note that only the measured resource states are displayed here. (e) The syndrome graph representation of (d) in the \emph{minimal} architecture: The edges are labelled by their corresponding $ZZ$-type measurement outcomes. The vertices are the parity check operators supported by the labels of their incident edges. (f) The syndrome graph representation of (d) in the \emph{cyclic} architecture: The internal vertex represents the parity check operator that is the product of the $ZZ$-type outcomes represented by the incident edges. The other vertices are the parity check operators associated with unit cells.}
\label{fig:QEC}

\end{figure*}

Let us describe the parity check operators and syndrome graph of the minimal architecture. For each cubic unit cell, which is shown in Fig.~\ref{fig:unit_cell}, of a RHG lattice, we define a parity check operator as $P_c = \prod_{f,e \in c} X_f Z_e$, where $c$ labels a cubic unit cell, $X_f$ and $Z_e$ act on a face and edge qubit, respectively. Note that there are four face qubits per face and two edge qubits per edge on a unit cell $c$. We now verify that $P_c \in \mathcal{C} = \mathcal{R}\cap\mathcal{M}$. $P_c$ is a product of 24 stabilizers, each from a different resource state on a unit cell, of the form $X_f Z_e$, as illustrated in Fig.~\ref{fig:QEC}(a); thus, $P_c \in \mathcal{R}$. As shown in Fig.~\ref{fig:QEC}(b), the product of the $\prod_i X_i$-type outcomes of the GSMs at the faces yields $\prod_{f} X_f$. For every edge $e \in c$, the two factors of $Z_e$ comes either directly as a $ZZ$-type outcome, or as a product of all $ZZ$-type outcomes of a minimal GSM. Therefore, $P_c \in \mathcal{M}$ and thus, $P_c \in \mathcal{C}$.
In order to construct the syndrome graph, we need to show explicitly how each $\prod_i X_i$- or $ZZ$-type GSM outcome supports two parity check operators (in the bulk). Since neighboring parity check operators always share a face, the $\prod_i X_i$-type outcome from the GSM at the face supports both operators. The example in Fig.~\ref{fig:QEC}(d) shows that a $ZZ$-type outcome from the GSM at an edge supports two parity check operators. Thus far, we have yet to consider the $ZZ$ and $\prod_i X_i$ GSM outcomes at the faces and edges, respectively. To incorporate them, we consider a shifted lattice where the edge (face) qubits are the face (edge) qubits in the original lattice; the original and shifted lattices are commonly known as the primal and dual lattices in the literature~\cite{raussendorf2007topological,fowler2009topological,bartolucci2023fusion}. Then, we can form the parity check operators and syndrome graph for the dual lattice from the $ZZ$- and $\prod_i X_i$-type GSM outcomes at the primal faces and edges, respectively, the same way we construct them for the primal lattice.

We now proceed to discuss the cyclic architecture. As in the minimal architecture, we define the parity check operator $P_c$ for each cubic unit cell. One can show $P_c \in \mathcal{C}$ for the same reasons in the minimal architecture, except here, a cyclic GSM at an edge will always directly impart the two factors of $Z_e$ per $ZZ$-type outcome to a $P_c$. Consequently, the syndrome graphs of the cyclic and minimal architectures will be different. This is so because while two $P_c$'s that share a face are supported by the same $\prod_i X_i$-type outcome, no two $P_c$'s share a $ZZ$-type outcome. In order to detect and correct errors in $ZZ$-type outcomes, we introduce an additional type of parity check operators, thereby increasing the number of vertices in the syndrome graph and enlarging the check group $\mathcal{C}$. Specifically, for each $n$-qubit GSM at an edge $e$, this parity check operator is defined as $P_e = \prod_{i=1}^n Z_{ia}Z_{ib}$, which is the product of all $ZZ$-type outcomes from the GSM (see Sec.~\ref{sec:encoded_GHZ}), implying that $P_e \in \mathcal{M}$. Since for all $i$, $Z_{ia}Z_{ib} \in \mathcal{R}$ (see Sec.~\ref{sec:resource}), $P_e \in \mathcal{R}$ and thus $\mathcal{C}$. Note that $P_e$ and $P_c$ cannot be generated from each other because any product of $P_c$'s contains factors of $X$, while $P_e$ does not. Moreover, a $ZZ$-type outcome at an edge $e$ will always support one $P_e$ and $P_c$ if $e$ is an edge of the unit cell $c$. As in the minimal architecture, the primal and dual lattices can be considered separately. By considering the dual lattice, we can account for the $ZZ$- and $\prod_i X_i$-type GSM outcomes at the primal faces and edges, respectively.

\section{Linear-Optical Architecture performance}
\label{sec:performance}

In this section, we describe our linear-optical implementations of the architectures, and evaluate their performance. We consider photonic dual-rail qubits~\cite{linear2007kok}, of which the computational basis states are encoded in a single photon that exists in one of two orthogonal modes, e.g., two spatial modes specified by distinct waveguides. As in Refs.~\cite{bartolucci2023fusion,sahay2023tailoring}, we adopt a linear-optical error model, where there are two sources of errors: photon loss and the probabilistic nature of BSMs between dual-rail qubits, both of which can cause erasure errors, as will be discussed below. The model further assumes that the loss, which is independent on all photons, occurs after an ideal resource state generation. This model can be translated into a more realistic model where the state generation is lossy, under assumptions about the distribution and correlation of loss across different components~\cite{varnava2008how,Brod2020classicalsimulation}.

In the absence of photon loss, a dual-rail BSM will probabilistically return the measurements of either (1) $X_1 X_2$ and $Z_1 Z_2$, or (2) $P_1$ and $P_2$, where $P \in \{ X, Z\}$ depends on the linear-optical circuit configuration of the BSM. The BSM circuits corresponding to different guaranteed bases $P$ differ only by single-qubit gates that can be easily implemented deterministically in linear optics. Scenario (1) is considered as a successful BSM, since it returns both desired two-qubit outcomes. In scenario (2), we can salvage a two-qubit measurement outcome $P_1 P_2$ by multiplying $P_1$ and $P_2$, but the other desired two-qubit outcome will be erased. Importantly, the two scenarios are indicated by distinct detector click patterns. Therefore, the erasure of a two-qubit outcome will always be heralded. In the presence of loss, the detectors will return fewer than two clicks, while an ideal BSM, which admits two photons and preserves the number of photons, always returns two clicks. This heralds a third scenario -- the erasure of both $X_1 X_2$ and $Z_1 Z_2$. A lossless BSM between two dual-rail qubits succeeds with probability $\frac{1}{2}$~\cite{calsamiglia2001maximum}. Assuming a photon is lost during a BSM with probability $\eta$, commonly referred to as \emph{single-photon loss rate}, the BSM success probability drops to $\frac{(1-\eta)^2}{2}$.

In our constructions, we enhance the success probability of each BSM under photon loss by encoding it in a small code\footnote{We note that alternative to applying an encoding, we could boost the success probability of a BSM by coupling it to ancillary photons~\cite{grice2011arbitrarily,ewert201434}. However, the ancillary photons are susceptible to loss and thus, increase the likelihood of an erasure. Furthermore, boosted BSMs have more stringent hardware requirements in practice~\cite{wein2016efficiency}; for instance, the detectors must resolve larger numbers of photons, and the ancillary photons could be entangled. Therefore, we choose not to adopt this method.}. The encoding, in turn, suppresses the probability that a GSM outcome is erased. Specifically, we will construct our architectures using encoded BSMs from Refs.~\cite{ewert2016ultrafast,lee2019fundamental} (see App.~\ref{app:QPC_BSM} for details.). Both schemes employ the quantum parity code (QPC) -- a concatenation of two repetition codes and a generalization of the Shor code~\cite{bacon2006quantum}, and implement an encoded BSM transversally as multiple dual-rail BSMs. Compared to the method in Ref.~\cite{ewert2016ultrafast}, which only requires static linear-optical components to realize, the one in Ref.~\cite{lee2019fundamental} achieves a higher success probability at the same loss rate and code size, at the cost of additional active optical components to perform feed-forward operations. Let $m_1$ and $m_2$ be the sizes of the two repetition codes in a QPC. It was shown in Refs. ~\cite{ewert2016ultrafast,lee2019fundamental} that given a single-photon loss rate, the BSM success probability can be optimized by tuning $m_1$ and $m_2$, as well as an additional feed-forward parameter $j$ for the encoded BSM in Ref.~\cite{lee2019fundamental}. The encoding requires replacing every qubit in a resource state by an encoded qubit, which contains $m_1 m_2$ qubits. As such, we can define two families of encoded resource states, which contain $3m_1 m_2$ and $4m_1 m_2$ dual-rail qubits, for the minimal and cyclic architectures, respectively. 

We numerically simulate the single-photon loss threshold for every construction. We briefly describe the numerical simulations and leave further details in App.~\ref{app:sim}. First, we derive analytical formulas, in terms of $\eta, m_1, m_2$ and $j$ when applicable, for the erasure probabilities of the $\prod_i X_i$- and $ZZ$-type GSM outcomes. We then perform Monte-Carlo simulations, in which, according to the analytical formulas, errors are applied to syndrome graphs. The syndrome graphs are derived from surface codes with boundaries, where logical errors manifest as connected chains of errors that span between two opposite boundaries. We evaluate the logical error rates at various surface-code distances, $\eta, m_1$ and $m_2$, as well as $j$ when applicable, and estimate the single-photon loss thresholds. 

\begin{figure}
    \centering
    \includegraphics[width = \columnwidth]{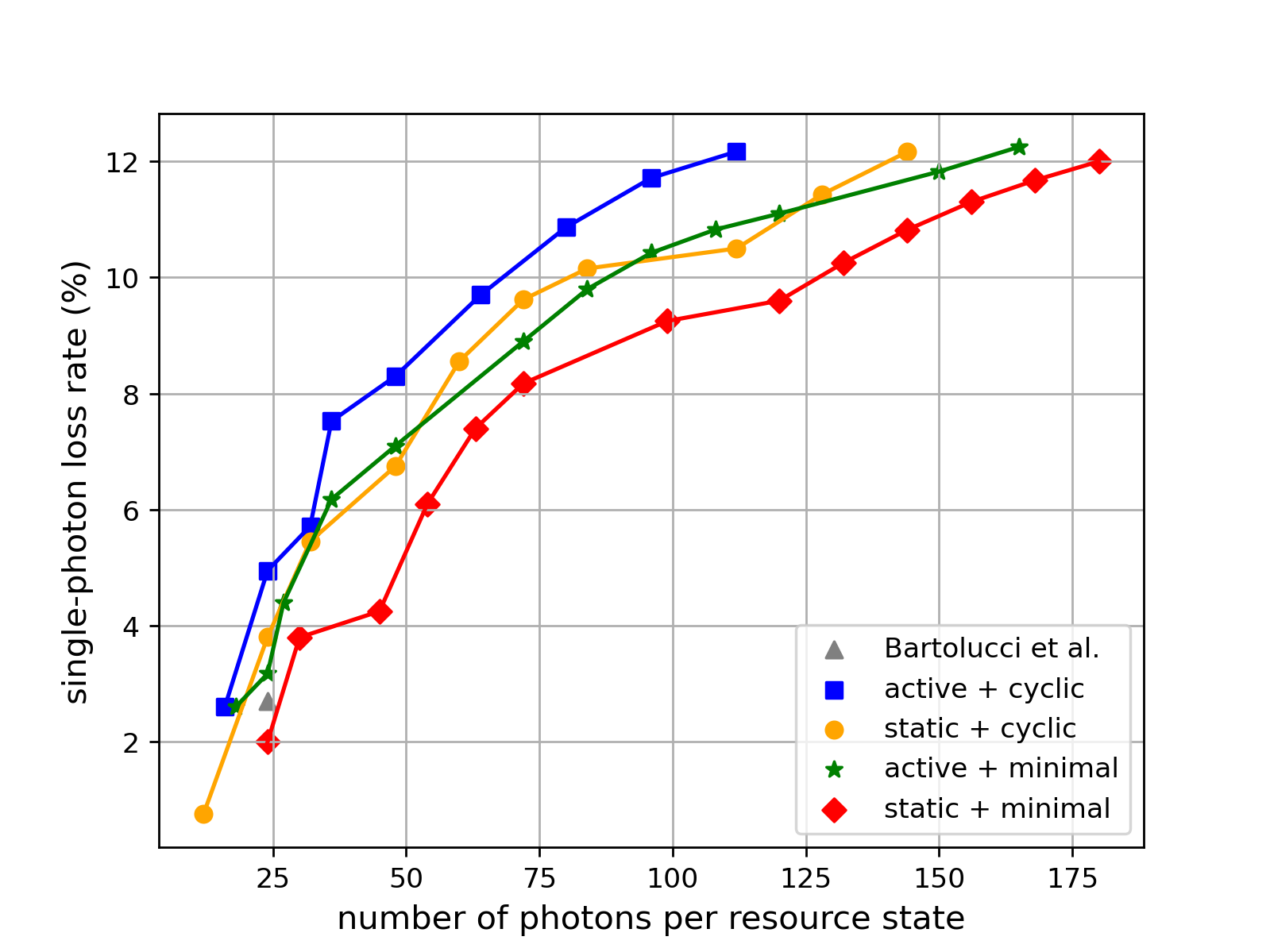}
    \caption{Single-photon loss thresholds for the minimal and cyclic architectures constructed using GHZ-state measurements and resource states that are encoded in a quantum parity code (QPC). The hardware implementation of the encoded measurements comprises either only static, or active linear-optical components. Each solid line shows thresholds for a choice of architecture, i.e., minimal or cyclic, and measurement implementation, i.e., using static or active linear optics, at various QPC code sizes and thus, number of photons per resource state. For comparison, we also show the threshold (gray triangle) of the construction from Ref.~\cite{bartolucci2023fusion} using a 24-photon QPC-encoded resource state.}
    \label{fig:threshold}
\end{figure}

We compare different constructions in Fig.~\ref{fig:threshold}, where their thresholds are plotted against the number of photons in their resource states. We observe that when constructed with resource states of a similar size and the same type of encoded BSM, the cyclic architecture always outperforms the minimal architecture.  We compare our cyclic architecture with the 24-photon, encoded six-ring FBQC construction, of which the single-photon loss threshold is $2.7\%$~\cite{bartolucci2021creation}, and find two significant improvements in that we obtain (1) a $75\%$ higher threshold $\sim 4.75\%$ using resource states of the same size, and (2) a similar threshold $\sim 2.5 \%$ using resource states with $\frac{1}{3}$ fewer photons, i.e., 16 photons. We further observe that for any combination of architecture, i.e., minimal or cyclic, and encoded BSM, i.e., static or active, the loss threshold can be improved by enlarging the encoding and thus, resource states. We include thresholds up to $\sim 12\%$ in Fig.~\ref{fig:threshold}, though higher thresholds are attainable, because the required number of photons per resource states far exceeds current experimental capabilities; the resource states that enable a $\sim 12\%$ threshold are roughly ten times larger than the largest, to our knowledge, experimentally generated photonic resource state~\cite{thomas2022efficient}.

\section{Discussion}
\label{sec:discussion}

In this work, we investigate MBQC architectures that achieve fault-tolerance by fusing two-qubit resource states via GHZ-state measurements. Tailoring the architectures to linear optics, we encode the GHZ-state measurements and resource states in a quantum parity code to suppress errors due to photon loss and the probabilistic nature of linear optics. Our simulations show that enlarging the quantum parity code and thus the resource states can increase photon-loss thresholds, thereby demonstrating the efficacy of our encoded GHZ-state measurements. Furthermore, we obtain high thresholds using resource states with modest numbers of photons, comparable to the largest experimentally generated photonic resource states.

A non-limiting list of exploration beyond our work shown herein includes encoding GHZ-state measurements in different codes~\cite{schmidt2019efficiencies,bell2022optimising}, exploiting biased errors~\cite{sahay2023tailoring}, employing different resource states~\cite{paesani2022high}, and considering fusion networks that are not derived from foliation~\cite{nickerson2018measurement,Newman2020generatingfault}. Beyond quantum computation, our encoded GHZ-state measurements may also find use in quantum communication (see App.~\ref{app:qcomm} for further discussion). For instance, they may be used to perform multipath routing, which can enhance entanglement rates in quantum networks~\cite{pirandola2019end,patil2022entanglement}.

\section*{Acknowledgements}
The authors thank Ian Walmsley, Josh Nunn, Alex Jones, Richard Tatham and our colleagues at ORCA computing for their helpful comments.


\bibliography{Papers}
\bibliographystyle{apsrev4-2}

\appendix

\section{Encoded Bell-state measurement}
\label{app:QPC_BSM}

Here we will summarize the static and active schemes, proposed respectively in Ref.~\cite{ewert2016ultrafast} and~\cite{lee2019fundamental}, for realizing a Bell-state measurement (BSM) encoded in a quantum parity code (QPC).
The physical measurements that constitute the encoded measurements are the \emph{dual-rail} BSMs.
These optical circuits project onto two product states and two out of the four Bell states, which are defined by 
\begin{equation}
    \bellstate{k}{l} = \frac{1}{\sqrt{2}}\left(\ket{0}\ket{k} + (-1)^l \ket{1}\ket{1-k}\right),
\end{equation}
where $k,l \in \{0,1\}$ are referred to as the parity and phase bits respectively. Note that the Bell states are eigenstates of the operators $XX$ and $ZZ$, of which the respective eigenvalues $zz, xx \in \{\pm 1 \}$ are related to $k$ and $l$ via $zz=(-1)^{k}$ and $xx=(-1)^{l}$. Furthermore, the two Bell states that the measurement circuit projects onto are guaranteed to share an eigenvalue, which implies four distinct BSMs, which we denote as $BSM_{zz/xx = \pm 1}$, where the subscript indicates the shared eigenvalue.

Now consider the repetition code, $\ket{0^{(m)}} := \ket{0}^{\otimes m}$ and $\ket{1^{(m)}} := \ket{1}^{\otimes m}$, where we have used the superscript $(m)$ to indicate a repetition encoded state.
The repetition-encoded Bell states can be decomposed in terms of those on the physical qubits as
\begin{subequations}\label{eq:mBell}
\begin{equation}
\mbellstate{0}{0}{(m)} = \bigg( \frac{1}{\sqrt{2}} \bigg)^{m-1} \sum_{r:\mathrm{even}}^m \mathcal{P}\left[ \bellstate{0}{0}^{\otimes m-r} \bellstate{0}{1}^{\otimes r} \right],
\end{equation}
\begin{equation}
\mbellstate{0}{1}{(m)} = \bigg( \frac{1}{\sqrt{2}} \bigg)^{m-1} \sum_{r:\mathrm{odd}}^m \mathcal{P}\left[ \bellstate{0}{0}^{\otimes m-r} \bellstate{0}{1}^{\otimes r} \right],
\end{equation}
\begin{equation}
\mbellstate{1}{0}{(m)} = \bigg( \frac{1}{\sqrt{2}} \bigg)^{m-1} \sum_{r:\mathrm{even}}^m \mathcal{P}\left[ \bellstate{1}{0}^{\otimes m-r} \bellstate{1}{1}^{\otimes r} \right],
\end{equation}
\begin{equation}
\mbellstate{1}{1}{(m)} = \bigg( \frac{1}{\sqrt{2}} \bigg)^{m-1} \sum_{r:\mathrm{odd}}^m \mathcal{P}\left[ \bellstate{1}{0}^{\otimes m-r} \bellstate{1}{1}^{\otimes r} \right],
\end{equation}
\end{subequations}
where $\mathcal{P}\left[\cdot \right]$ takes a tensor product of $m$ Bell states and outputs a sum over all permutations involving the $m$ tensor factors.

A QPC is a concatenation of two repetition codes.
In particular, we will adopt the QPC encoding convention used in~\cite{ewert2016ultrafast}, where
\begin{equation}\label{eq:qpc_had}
 \ket{\pm^{(n,m)}} := \ket{\pm^{(m)}}^{\otimes n} = \left[\frac{1}{\sqrt{2}} \left(\ket{0}^{\otimes m} \pm \ket{1}^{\otimes m} \right) \right]^{\otimes n}.
\end{equation}
We have used and will hereafter use the superscript $(n,m)$ to denote a QPC-encoded state or operator.
Then, the QPC-encoded Bell states admit the following decompositions~\cite{ewert2016ultrafast,lee2019fundamental}
 \begin{subequations}\label{eq:qpcBell}
 \begin{equation}
 \mbellstate{0}{0}{(n,m)} = \bigg( \frac{1}{\sqrt{2}} \bigg)^{n-1} \sum_{r:\mathrm{even}}^n \mathcal{P}\left[ \mbellstate{0}{0}{(m)}^{\otimes n-r} \mbellstate{1}{0}{(m)}^{\otimes r} \right],
 \label{eq:qpcBella}
 \end{equation}
 \begin{equation}
 \mbellstate{0}{1}{(n,m)} = \bigg( \frac{1}{\sqrt{2}} \bigg)^{n-1} \sum_{r:\mathrm{even}}^n \mathcal{P}\left[ \mbellstate{0}{1}{(m)}^{\otimes n-r} \mbellstate{1}{1}{(m)}^{\otimes r} \right],
 \label{eq:qpcBellb}
 \end{equation}
 \begin{equation}
 \mbellstate{1}{0}{(n,m)} = \bigg( \frac{1}{\sqrt{2}} \bigg)^{n-1} \sum_{r:\mathrm{odd}}^n \mathcal{P}\left[ \mbellstate{0}{0}{(m)}^{\otimes n-r} \mbellstate{1}{0}{(m)}^{\otimes r} \right],
 \label{eq:qpcBellc}
 \end{equation}
 \begin{equation}
 \mbellstate{1}{1}{(n,m)} = \bigg( \frac{1}{\sqrt{2}} \bigg)^{n-1} \sum_{r:\mathrm{odd}}^n \mathcal{P}\left[ \mbellstate{0}{1}{(m)}^{\otimes n-r} \mbellstate{1}{1}{(m)}^{\otimes r} \right].
 \label{eq:qpcBelld}
 \end{equation}
 \end{subequations}

From the Eqns. (\ref{eq:qpcBell}a-d) we see that we need only to extract the $XX^{(m)}$ eigenvalue from any one of the $n$ blocks to infer the $XX^{(n,m)}$ eigenvalue.
On the other hand, we need to extract the $ZZ^{(m)}$ eigenvalue from all $n$ blocks in order to infer that of $ZZ^{(n,m)}$.
Let $P(zz^{(n,m)})$ and $P(xx^{(n,m)})$ denote the probabilities of obtaining the eigenvalues of the QPC-encoded operators $ZZ^{(n,m)}$ and $XX^{(n,m)}$ respectively, while $P(zz^{(m)})$ and $P(xx^{(m)})$ denote those of the block-level operators $ZZ^{(m)}$ and $XX^{(m)}$, respectively.
The probabilities for the QPC-encoded operators can then generally be related to those for the block-level operators by
\begin{align}\label{eq:zzSuccess}
 & P(zz^{(n,m)}) = \big(P(zz^{(m)}) \big)^n, \\
 \label{eq:xxSuccess} & P(xx^{(n,m)}) = 1 - \big(1 - P(xx^{(m)}) \big)^n.
\end{align}

Note that if one chooses to adopt the Shor-style QPC,
\begin{equation}\label{eq:qpc_shor}
    \ket{0^{(n,m)}} := \ket{+^{(m)}}^{\otimes n}, \: \ket{1^{(n,m)}} := \ket{-^{(m)}}^{\otimes n}
\end{equation}
as in~\cite{lee2015nearly_prl}, the results of this section can be adapted by replacing $zz^{(n,m)}$ with $xx^{(n,m)}$, and vice versa.

In the following subsections we will consider the block-level measurement schemes proposed in Ref.~\cite{ewert2016ultrafast} and~\cite{lee2019fundamental}, respectively, with a focus on their probabilities for producing the individual Bell state eigenvalues.

\subsection{Ewert-Bergmann-van-Loock Protocol}
\label{sec:EBV_BSM}

Given Bell states encoded as in Eqn.~\eqref{eq:mBell}, one performs $m$ dual-rail BSMs, e.g., $BSM_{zz=-1}$, in parallel, i.e., one on each factor of $\bellstate{k}{l}$, according to the Ewert-Bergmann-van-Loock protocol.
We will refer to this block-level measurement scheme as $BSM_{zz^{(m)} = -1}$.
In what follows we will use $\gamma$ to denote the probability that a dual-rail BSM is corrupted and does not return any information.

From Eqn.~\eqref{eq:mBell} we see that when we apply $BSM_{zz^{(m)} = -1}$ to any state $\mbellstate{k}{l}{(m)}$ then we are able to infer the $ZZ^{(m)}$ eigenvalue as long as at least one dual-rail BSM returns the $ZZ$ eigenvalue.
On the other hand, the extraction of the $XX^{(m)}$ eigenvalue depends on the input state $\mbellstate{k}{l}{(m)}$.
If $k=0$, then from Eqns.~\eqref{eq:mBell} we see that there is no way to infer the $XX^{(m)}$ eigenvalue for such a state using the dual-rail Bell state measurement $BSM_{zz=-1}$ (since it only projects onto $\bellstate{1}{0}$ and $\bellstate{1}{1}$).
Conversely, if $k=1$, then as long as no dual-rail BSM is corrupted, we are able to infer both $xx^{(m)}$ and $zz^{(m)}$.
Since the states with $k=0$ and $k=1$ occur with equal weight in the expansions~\eqref{eq:qpcBell} we arrive at the following rates
\begin{align}
	& P(zz^{(m)}) = 1 - \gamma^m, \\
	& P(xx^{(m)},zz^{(m)} ) = \frac{1}{2}(1-\gamma)^m.
\end{align}
Note that there is no scenario under which we obtain $xx^{(m)}$ but not $zz^{(m)}$ using this block-level measurement.

In Ref.~\cite{ewert2016ultrafast,ewert2017ultrafast} this $BSM_{zz^{(m)}=-1}$ protocol is applied to each block in Eqn.~\eqref{eq:qpcBell} to realize a Bell state measurement on QPC-encoded states.
It was shown~\cite{ewert2016ultrafast,ewert2017ultrafast} that the efficiency of a QPC Bell state measurement, averaged over all four QPC-encoded Bell states, is
\begin{equation}\label{eq:EBL_eff}
P\big(xx^{(n,m)}, zz^{(n,m)}\big) = \big(1 - \gamma^m\big)^n - \big(1 - \gamma^m - \frac{1}{2}(1-\gamma)^m\big)^n.
\end{equation}
Using Eqns.~\eqref{eq:zzSuccess} and~\eqref{eq:xxSuccess} we find the rates for extracting the individual eigenvalues with this scheme to be
\begin{align}
& P\big(zz^{(n,m)} \big) = \big(1-\gamma^m \big)^n, \label{eq:EBL_zprob}\\
& P\big(xx^{(n,m)} \big) = 1 - \big(1-\frac{1}{2}(1-\gamma)^m \big)^n. \label{eq:EBL_xprob}
\end{align}

A dual-rail BSM is corrupted unless both input photons are detected. Let $\eta$ be the single-photon loss rate or loss rate per photon.
Then, $\gamma = 1-(1-\eta)^2$.
We could alternatively consider a case where only one photon entering a dual-rail BSM is subjected to loss by simply setting $\gamma = \eta$ to recover the efficiency as derived in \cite{ewert2016ultrafast,ewert2017ultrafast}.
Further generalizations to where both photons are subject to loss, but at different rates, is straightforward.

\subsection{Lee-Ralph-Jeong Protocol}
\label{sec:LRJ_BSM}

The QPC-encoded BSM in ref.~\cite{lee2015nearly_prl} is a protocol that features an adaptive block-level measurement scheme, denoted $BSM_{(1)}$.
In what follows we will be using $\gamma$ to carry the same meaning as in the previous subsection. 

We begin by considering the block-level Bell states in Eqn.~\eqref{eq:mBell}.
As before, we will perform $m$ dual-rail BSMs, one on each factor of $\bellstate{k}{l}$, but in this scheme we will be using the dual-rail measurements $BSM_{xx=-1}$ as well as $BSM_{zz=\pm1}$.
Let $0 < j < m$ (the value of $j$ will be numerically optimized) and $1\leq i<j$.
We select $j$ arbitrary factors of $\bellstate{k}{l}$, and perform $BSM_{xx=-1}$ on the $i+1$-th factor if the $i$-th $BSM_{xx=-1}$ has failed to return $\bellstate{0}{1}$ or $\bellstate{1}{1}$. If the $i$-th $BSM_{xx=-1}$ succeeds and returns $\bellstate{0}{1}$ ($\bellstate{1}{1}$), $BSM_{zz=+1}$ ($BSM_{zz=-1}$) will be applied to all the remaining $m-i$ factors of $\bellstate{k}{l}$. If a photon loss is detected at the $i$-th $BSM_{xx=-1}$ or if $BSM_{xx=-1}$ fails $j$ times in a row, one of $BSM_{zz=\pm 1}$ is selected at random to be applied to the remaining factors of $\bellstate{k}{l}$. From the $m$ BSMs, we can resolve either (i) both the $xx^{(m)}$ and $zz^{(m)}$ eigenvalues, (ii) just the $zz^{(m)}$ eigenvalues, or (iii) no eigenvalues.

 The adaptive protocol $BSM_{(1)}$ does not have the same degree of sensitivity to the state of the input block as seen with the scheme in the previous subsection.
In order to succeed it is required that all dual-rail BSMs are executed free of photon loss.
Additionally, we need either (1) one of the $BSM_{xx=-1}$'s to succeed, or (2) in the event all $j$ of the $BSM_{xx=-1}$'s fail, to guess right in our choice of $BSM_{zz=\pm 1}$.
 This leads to a efficiency for the $BSM_{(1)}$ scheme of
  \begin{equation}
     P(zz^{(m)},xx^{(m)}) = \Big(1- \frac{1}{2^{j+1}} \Big) (1-\gamma)^{m}.
 \end{equation}

Photon loss can lead to the corruption of a block-level BSM, i.e., revealing neither $xx^{(m)}$ nor $zz^{(m)}$.
This happens when $l \leq j$ consecutive $BSM_{xx=-1}$'s fail followed by corruption of the $m-l$ remaining dual-rail BSMs.
Thus, the scheme $BSM_{(1)}$ returns no information with probability
\begin{equation}
 P(\no{zz^{(m)}}, \no{xx^{(m)}}) = \sum_{l=0}^{j} \left(\frac{1-\gamma}{2}\right)^{l} \gamma^{m-l}.
\end{equation}
Since there is no scenario in which $xx^{(m)}$ is the sole bit of information returned, the probability of obtaining $zz^{(m)}$ must be
 \begin{equation}
     P(zz^{(m)}) = 1- \sum_{l=0}^{j} \left(\frac{1-\gamma}{2}\right)^{l} \gamma^{m-l}.
 \end{equation}

In Ref.~\cite{lee2015nearly_prl} it was shown that the when applying the adaptive measurement scheme $BSM_{(1)}$ to the blocks in Eqn.~\eqref{eq:qpcBell} the efficiency of the QPC Bell state measurement is
\begin{multline}\label{eq:LRJ_eff}
	P\big( xx^{(n,m)}, zz^{(n,m)} \big) = \big( P(zz^{(m)})  \big)^n \\ - \big(P(zz^{(m)})  - P(zz^{(m)}, xx^{(m)}) \big)^n.
\end{multline}
Furthermore, from Eqns.~\eqref{eq:zzSuccess} and~\eqref{eq:xxSuccess} we compute the probabilities for obtaining the individual eigenvalues of the operators $ZZ^{(n,m)}$ and $XX^{(n,m)}$ to be
\begin{align}
 & P\big(zz^{(n,m)}\big) = \bigg(1- \sum_{l=0}^{j} \left(\frac{1-\gamma}{2}\right)^{l} \gamma^{m-l}\bigg)^n,\label{eq:LRJ_zprob} \\
 & P\big(xx^{(n,m)}\big) = 1 - \bigg(1 - \Big(1-\frac{1}{2^{j+1}}\Big)\big(1-\gamma\big)^m\bigg)^n.\label{eq:LRJ_xprob}
\end{align}
Note that these expressions reduce to those in Eqns.~\eqref{eq:EBL_zprob} and~\eqref{eq:EBL_xprob} when $j=0$. In other words, if the feed-forward operations are not allowed, i.e., $j=0$, in the active encoded BSM, it is the same as static encoded BSM.

\section{Details on encoded GHZ-state measurements}
\label{app:ghz}

\subsection{Erasure probabilities}
\label{app:ghz_erasure}
We analytically derive the erasure probability of each outcome in a GHZ-state measurement (GSM), where its constituent BSMs are encoded according to Refs.~\cite{ewert2016ultrafast,lee2019fundamental} and delineated in App.~\ref{app:QPC_BSM}.

\textbf{Minimal GSM:}
An $n$-qubit minimal GSM is constructed from $n-1$ BSM. The GSM outcome $\prod_{i=1}^n X_i$ is obtained as a product of the $n-1$ $XX$-type BSM outcome, where as a $ZZ$-type GSM outcome is taken directly from a BSM. Therefore, the probability of obtaining a $\prod_{i=1}^n X_i$ and $ZZ$-type GSM outcome are
\begin{equation}\label{eq:erase_p1}
    p_{GSM}\left(\prod_{i=1}^n X_i\right) = \big(P(xx^{(n,m)})\big)^{n-1},
\end{equation}
and
\begin{equation}\label{eq:erase_p2}
    p_{GSM}(ZZ) = P(zz^{(n,m)}),
\end{equation}
respectively, where $P(xx^{(n,m)})$ and $P(zz^{(n,m)})$ respectively are the probabilities of successfully measuring the logical $XX$ and $ZZ$ in a QPC-encoded BSM, as defined in Eqns.~\eqref{eq:EBL_xprob} and~\eqref{eq:EBL_zprob} for the static BSM, and Eqns.~\eqref{eq:LRJ_xprob} and~\eqref{eq:LRJ_zprob} for the active BSM. The erasure probability of a $\prod_{i=1}^n X_i$ or $ZZ$-type GSM outcome is simply the complement of the probability of obtaining it.

\textbf{Cyclic GSM:}
An $n$-qubit minimal GSM is constructed from $n$ BSM. The GSM outcome $\prod_{i=1}^n X_i$ is obtained as a product of the $n$ $XX$-type BSM outcome, where as a $ZZ$-type GSM outcome is taken directly from a BSM. Similar to the minimal GSM, the probability of obtaining a $\prod_{i=1}^n X_i$ and $ZZ$-type GSM outcome are
\begin{equation}\label{eq:erase_p3}
    p_{GSM}\left(\prod_{i=1}^n X_i \right) = \big(P(xx^{(n,m)})\big)^{n},
\end{equation}
and
\begin{equation}\label{eq:erase_p4}
    p_{GSM}(ZZ) = P(zz^{(n,m)}),
\end{equation}
respectively. Once again, $P(xx^{(n,m)})$ and $P(zz^{(n,m)})$ respectively are the probabilities of successfully measuring the logical $XX$ and $ZZ$ in a QPC-encoded BSM, as defined in Eqns.~\eqref{eq:EBL_xprob} and~\eqref{eq:EBL_zprob} for the static BSM, and Eqns.~\eqref{eq:LRJ_xprob} and~\eqref{eq:LRJ_zprob} for the active BSM. The erasure probability of a $\prod_{i=1}^n X_i$ or $ZZ$-type GSM outcome is simply the complement of the probability of obtaining it.

\subsection{Efficiency under photon loss}
\label{app:qcomm}

In quantum communication, projective measurements onto entangled bases, e.g., Bell or GHZ basis, are used to distribute entanglement across a spatial network~\cite{ewert2016ultrafast,lee2019fundamental,pirandola2019end,patil2022entanglement,bell2022optimising}. In optical quantum networking protocols, it is paramount for the entangling measurements to attain high efficiencies under photon loss. Efficiency is defined as the probability that all the operators' eigenvalues in the desired entangled basis are extracted. For instance, the efficiency of a BSM is the probability that the eigenvalues of $XX$ and $ZZ$ are extracted, and that of a GSM is the probability that eigenvalues of the operators in $\{ \prod_{i=1}^n X_i , Z_{j}Z_{j+1} \}_{j=1}^{n-1}$ are extracted.

Here we derive the efficiencies of our encoded minimal and cyclic GSM, which are constructed from QPC-encoded BSMs~\cite{ewert2016ultrafast,lee2019fundamental}. In order to extract all the relevant eigenvalues in a minimal GSM, we need all $n-1$ constituent encoded BSMs to successfully yield both $xx^{(n,m)}$ and $zz^{(n,m)}$, two-qubit Pauli observables encoded in a QPC and defined in App.~\ref{app:QPC_BSM}. As such, the encoded minimal GSM efficiency is
\begin{equation}\label{eq:mGSM_eff}
    e_{min} = \big(P( xx^{(n,m)}, zz^{(n,m)})\big)^{n-1},
\end{equation}
where $P( xx^{(n,m)}, zz^{(n,m)})$ is the efficiency of a QPC-encoded BSM, defined in Eqns.~\eqref{eq:EBL_eff} and~\eqref{eq:LRJ_eff} for the static and active encoded BSMs respectively. In the case of a cyclic GSM, we need all $n$ constituent encoded BSMs to return $xx^{(n,m)}$ in order to reconstruct the logical $\prod_{i=1}^n X_i$; we only need $n-1$ of the BSMs to return $zz^{(n,m)}$ to deduce the logical $Z_{j}Z_{j+1}$'s. As such, the encoded cyclic GSM efficiency is
\begin{multline}\label{eq:cGSM_eff}
    e_{cyc} = \big(P( xx^{(n,m)}, zz^{(n,m)} )\big)^{n} + \\ n \times\big(P( xx^{(n,m)}, zz^{(n,m)} )\big)^{n-1} P(xx^{(n,m)},\no{{zz^{(n,m)}}}),
\end{multline}
where
\begin{equation}
    P(xx^{(n,m)},\no{zz^{(n,m)}}) = P(xx^{(n,m)}) - P\big( xx^{(n,m)}, zz^{(n,m)} \big).
\end{equation}

It was shown respectively in Ref.~\cite{ewert2016ultrafast} and~\cite{lee2019fundamental} that the static and active encoded BSM can achieve efficiencies arbitrarily close to unity over a range of single-photon loss rates by enlarging the size of the QPC-encoding. Therefore, in accordance with Eqns.~\eqref{eq:mGSM_eff} and~\eqref{eq:cGSM_eff}, so can our encoded minimal and cyclic GSM. We provide supporting numerical evidences in Tables~\ref{tb:X4GHZ}--\ref{tb:Z4GHZ-ff}, where we include the efficiencies of 4-qubit GSMs at various QPC code sizes and single-photon loss rates. Note that we use the QPC definition in Eqn.~\eqref{eq:qpc_shor}. 

We expect that our GSMs may be used to perform multipath routing, which can enhance entanglement rates in quantum networks~\cite{pirandola2019end,patil2022entanglement}. To this end, our protocols could potentially be used to form the basis of optical implementations of the networking protocols in Ref.~\cite{patil2022entanglement}, where probabilistic GHZ measurements are performed at repeater nodes. The analysis of such implementations under errors such as measurement failures and photon loss constitutes an intriguing direction for future work.

\begin{table*}[htbp!]
\begin{tabular}{l|l|l|l|l|l|l|l|l|l}
      & 0      & 0.001  & 0.01   & 0.02   & 0.03   & 0.04   & 0.05   & 0.08   & 0.1    \\ \hline
(2,2) & 0.7383 & 0.7349 &  &  &  &        &        &        &        \\ \hline
(2,3) & 0.7383 & 0.7332 &  &  &  &  &  &        &        \\ \hline
(3,1) & 0.9211 & 0.8993 & 0.7237 &        &        &        &        &        &        \\ \hline
(3,2) & 0.9211 & 0.9194 & 0.8987 & 0.8664 & 0.8256 & 0.7779 & 0.7247 &        &        \\ \hline
(3,3) & 0.9211    & 0.9185 & 0.8927 & 0.8587 & 0.8193 & 0.7748 & 0.7258   &  &        \\ \hline
(3,4) & 0.9211    & 0.9177    & 0.8822 & 0.8347 & 0.7797 & 0.7188 &  &  &        \\ \hline
(4,1) & 0.9785 & 0.9476 &  &        &        &        &        &        &        \\ \hline
(4,2) & 0.9785 & 0.9777 & 0.9650 & 0.9385 & 0.9002 & 0.8518 & 0.7952&        &        \\ \hline
(4,3) & 0.9785    & 0.9775 & 0.9667 & 0.9503 & 0.9282 & 0.8999 & 0.8647 & 0.7214 &        \\ \hline
(4,4) & 0.9785  & 0.9771    & 0.9621 & 0.9386 & 0.9070 & 0.8670 & 0.8189 & &  \\ \hline
(5,1) & 0.9944 & 0.9554 &        &        &        &        &        &        &        \\ \hline
(5,2) & 0.9944    & 0.9941 & 0.9840 & 0.9580 & 0.9178 & 0.8655 & 0.8035 &        &        \\ \hline
(5,3) & 0.9944    & 0.9940    & 0.9901 & 0.9827 & 0.9710 & 0.9535 & 0.9294 & 0.8120 &        \\ \hline
(5,4) & 0.9944    & 0.9939    & 0.9884 & 0.9783 & 0.9627 & 0.9402 & 0.9098 & 0.7681  &  \\ \hline
(6,1) & 0.9986 & 0.9517 &        &        &        &        &        &        &        \\ \hline
(6,2) & 0.9986    & 0.9984 & 0.9883 & 0.9597 & 0.9148 & 0.8563 & 0.7873 &        &        \\ \hline
(6,3) & 0.9986    & 0.9985    & 0.9970 & 0.9934 & 0.9863 & 0.9742 & 0.9561 & 0.8561 & 0.750  \\ \hline
(6,4) & 0.9986    & 0.9984    & 0.9965 & 0.9925 & 0.9854 & 0.9736 & 0.9558 & 0.8539 & 0.7429 \\ \hline
(7,1) & 0.9996 & 0.9452 &        &        &        &        &        &        &        \\ \hline
(7,2) & 0.9996    & 0.9995 & 0.9884 & 0.9560 & 0.9052 & 0.8394 & 0.7627 &        &        \\ \hline
(7,3) & 0.9996    & 0.9996 & 0.9990 & 0.9966 & 0.9912 & 0.9812 & 0.9654 & 0.8731 & 0.7706 \\ \hline
(7,4) & 0.9996    & 0.9996    & 0.9990 & 0.9974 & 0.9942 & 0.9882 & 0.9779 & 0.9066 & 0.8147
\end{tabular}
\caption{\textbf{4-qubit static, cyclic GSM efficiencies.} The constituent BSMs are implemented using the static, QPC-encoded BSM in Ref.~\cite{ewert2016ultrafast}. Here, $n,m$, shown in the first column, are the considered sizes of the two repetition codes in a QPC. We assume that each photon partaking in a BSM undergoes a loss at rates listed in the first row. The efficiencies far below 0.75 are omitted.}
\label{tb:X4GHZ}
\end{table*}

\begin{table*}[htbp!]
\begin{tabular}{l|l|l|l|l|l|l|l|l|l}
      & 0      & 0.001  & 0.01   & 0.02   & 0.03   & 0.04   & 0.05   & 0.08   & 0.1    \\ \hline
(4,1) & 0.8240 & 0.8044 &  &        &        &        &        &        &        \\ \hline
(4,2) & 0.8240 & 0.8213 & 0.7932 & 0.7546 &  &  & &        &        \\ \hline
(4,3) & 0.8240    & 0.8200 & 0.7825 & 0.7375 &  &  &  &  &        \\ \hline
(4,4) & 0.8240  & 0.8187    & 0.7681 & 0.7073 &  &  &  & &  \\ \hline
(5,1) & 0.9091 & 0.8823 &        &        &        &        &        &        &        \\ \hline
(5,2) & 0.9091    & 0.9073 & 0.8854 & 0.8504 & 0.8058 & 0.7533 & &        &        \\ \hline
(5,3) & 0.9091    & 0.9065    & 0.8804 & 0.8468 & 0.8081 & 0.7646 & 0.7167 &  &        \\ \hline
(5,4) & 0.9091    & 0.9056    & 0.8701 & 0.8236 & 0.7705 & 0.7122 & &   &  \\ \hline
(6,1) & 0.9539 & 0.9201 &        &        &        &        &        &        &        \\ \hline
(6,2) & 0.9539    & 0.9527 & 0.9355 & 0.9031 & 0.8583 & 0.8035 & 0.7411 &        &        \\ \hline
(6,3) & 0.9539    & 0.9522    &  0.9354 &  0.9120  &  0.8830  &  0.8480  &  0.8070  & &   \\ \hline
(6,4) & 0.9539    & 0.9516    & 0.9285 & 0.8958 & 0.8556 & 0.8083 & 0.7547 & &  \\ \hline
(7,1) & 0.9767 & 0.9366 &        &        &        &        &        &        &        \\ \hline
(7,2) & 0.9767    & 0.9760 & 0.9616 & 0.9302 & 0.8842 & 0.8264 & 0.7595 &        &        \\ \hline
(7,3) & 0.9767    & 0.9758 & 0.9653 & 0.9497 & 0.9285 & 0.9011 & 0.8668 & 0.7241 & \\ \hline
(7,4) & 0.9767    & 0.9754    & 0.9611 & 0.9392 & 0.9103 & 0.8740 & 0.8303 & & 
\end{tabular}
\caption{\textbf{4-qubit static, minimal GSM efficiencies.} The constituent BSMs are implemented using the static, QPC-encoded BSM in Ref.~\cite{ewert2016ultrafast}. Here, $n,m$, shown in the first column, are the considered sizes of the two repetition codes in a QPC. We assume that each photon partaking in a BSM undergoes a loss at rates listed in the first row. The efficiencies far below 0.75 are omitted.}
\label{tb:Z4GHZ}
\end{table*}

\begin{table*}[htbp!]
\begin{tabular}{l|l|l|l|l|l|l|l|l|l}
      & 0      & 0.001  & 0.01   & 0.02   & 0.03   & 0.04   & 0.05   & 0.08   & 0.1    \\ \hline
 (1,3)  &  0.9211  &  0.9137  &  0.8435  &   0.7613  & & & &  \\ \hline
 (1,4)  &  0.9785  &  0.9725  &  0.9031  &   0.8058  & & & &  \\ \hline
 (1,5)  &  0.9944  &  0.9901  &  0.9192  &   0.8015  & & & &  \\ \hline
(2,2) & 0.9785 & 0.9699 & 0.8942 & 0.8131 & 0.7356 &        &        &        &        \\ \hline
(2,3) & 0.9986 & 0.9944 & 0.9588 & 0.9291 & 0.8900 & 0.8423 & 0.7875 &        &        \\ \hline
(2,4) & 0.9999 & 0.9982 & 0.9914 &  0.9736 & 0.9426 & 0.8977  & 0.8400 &  & \\ \hline
 (2,5)  &  1.0  &  0.9998  &  0.9965  &   0.9819  &  0.9492 &  0.8960 &  0.8242 &  & \\ \hline
(3,1) & 0.9211 & 0.8993 & 0.7237 &        &        &        &        &        &        \\ \hline
(3,2) & 0.9986 & 0.9866 & 0.8987 & 0.8664 & 0.8256 & 0.7779 & 0.7248 &        &        \\ \hline
(3,3) & 1.0    & 0.9984 & 0.9939 & 0.9823 & 0.9630 & 0.9357 & 0.90   & 0.7481 &        \\ \hline
(3,4) & 1.0    & 1.0    & 0.9985 & 0.9937 & 0.9840 & 0.9678 & 0.9429 & 0.8071 &        \\ \hline
(4,1) & 0.9785 & 0.9476 & 0.7094 &        &        &        &        &        &        \\ \hline
(4,2) & 0.9999 & 0.9840 & 0.9650 & 0.9385 & 0.9002 & 0.8518 & 0.7952  &        &        \\ \hline
(4,3) & 1.0    & 0.9999 & 0.9965 & 0.9861 & 0.9686 & 0.9438 & 0.9116 & 0.7726 &        \\ \hline
(4,4) & 1.0    & 1.0    & 0.9995 & 0.9978 & 0.9935 & 0.9848 & 0.9698 & 0.8706 & 0.7611 \\ \hline
(5,1) & 0.9944 & 0.9554 &        &        &        &        &        &        &        \\ \hline
(5,2) & 1.0    & 0.9941 & 0.9840 & 0.9580 & 0.9178 & 0.8655 & 0.8035 &        &        \\ \hline
(5,3) & 1.0    & 1.0    & 0.9959 & 0.9837 & 0.9710 & 0.9535 & 0.9294 & 0.8120 &        \\ \hline
(5,4) & 1.0    & 1.0    & 0.9999 & 0.9991 & 0.9970 & 0.9924 & 0.9841 & 0.920  & 0.8311 \\ \hline
(5,5) & 1.0 & 1.0 & 1.0 & 0.9997 & 0.9988 & 0.9980 & 0.9938 & 0.9375 & 0.8427 \\ \hline
(6,1) & 0.9986 & 0.9517 &     &        &        &        &        &        &        \\ \hline
(6,2) & 1.0    & 0.9984 & 0.9883 & 0.9597 & 0.9149 & 0.8563 & 0.7873 &        &        \\ \hline
(6,3) & 1.0    & 1.0    & 0.9970 & 0.9934 & 0.9863 & 0.9743 & 0.9561 & 0.8561 & 0.750  \\ \hline
(6,4) & 1.0    & 1.0    & 0.9999 & 0.9992 & 0.9972 & 0.9933 & 0.9864 & 0.9357 & 0.8632 \\ \hline
(6,5) & 1.0 & 1.0 & 1.0 &  0.9999 & 0.9995 & 0.9980 & 0.9943 & 0.9596 & 0.8937 \\ \hline
(7,1) & 0.9996 & 0.9452 &        &        &        &        &        &        &        \\ \hline
(7,2) & 1.0    & 0.9995 & 0.9884 & 0.9561 & 0.9053 & 0.8394 & 0.7627 &        &        \\ \hline
(7,3) & 1.0    & 0.9999 & 0.9990 & 0.9966 & 0.9912 & 0.9812 & 0.9654 & 0.8731 & 0.7706 \\ \hline
(7,4) & 1.0    & 1.0    & 0.9999 & 0.9991 & 0.9969 & 0.9926 & 0.9855 & 0.9372 & 0.8715
\end{tabular}
\caption{\textbf{4-qubit active, cyclic GSM efficiencies.} The constituent BSMs are implemented using the active, QPC-encoded BSM in Ref.~\cite{lee2019fundamental}. Here, $n,m$, shown in the first column, are the considered sizes of the two repetition codes in a QPC. We assume that each photon partaking in a BSM undergoes a loss at rates listed in the first row. For every combination of $n,m$ and loss rate, we optimize the efficiency over $j$. The efficiencies far below 0.75 are omitted.}
\label{tb:X4GHZ-ff}
\end{table*}

\begin{table*}[htbp!]
\begin{tabular}{l|l|l|l|l|l|l|l|l|l}
      & 0      & 0.001  & 0.01   & 0.02   & 0.03   & 0.04   & 0.05   & 0.08   & 0.1    \\ \hline
 (1,4) &  0.8240 &  0.80442 &  &  & & & & &  \\ \hline
(2,2) & 0.8240 & 0.8161 & 0.7467 & &      &        &        &        &        \\ \hline
(2,3) & 0.9539 & 0.9475 & 0.8868 & 0.8148 & 0.7405 & &      &        &        \\ \hline
(3,2) & 0.9539 & 0.9441 & 0.8582 & 0.7679 &   &  &  &        &        \\ \hline
(3,3) & 0.9942    & 0.9890 & 0.9415 & 0.8862 & 0.8369 & 0.7822 & 0.7223   &  &        \\ \hline
(3,4) & 0.9993    & 0.9967    & 0.9779 & 0.9475 & 0.9031 & 0.8460 & 0.7790 &  &        \\ \hline
(4,1) & 0.8240 & 0.8044 &  &        &        &        &        &        &        \\ \hline
(4,2) & 0.9883 & 0.9761 & 0.8703 & 0.7619 &  &  &        &        &        \\ \hline
(4,3) & 0.9993    & 0.9932 & 0.9755 & 0.9533 & 0.9217 & 0.8810 & 0.8322 &  &        \\ \hline
(4,4) & 1.0    & 0.9991    & 0.9947 & 0.9827 & 0.9610 & 0.9285 & 0.8847 &  &  \\ \hline
(5,1) & 0.9091 & 0.8823 &       &        &        &        &        &        &        \\ \hline
(5,2) & 0.9971    & 0.9821 & 0.8854 & 0.8504 & 0.8058 & 0.7533 &  &        &        \\ \hline
(5,3) & 0.9999    & 0.9968    & 0.9904 & 0.9754 & 0.9514 & 0.9183 & 0.8766 &  &        \\ \hline
(5,4) & 1.0    & 0.9999   & 0.9977 & 0.9909 & 0.9779 & 0.9572 & 0.9275 & 0.7810  & \\ \hline
(6,1) & 0.9539 & 0.9201 &        &        &        &        &        &        &        \\ \hline
(6,2) & 0.9993    & 0.9814 & 0.9355 & 0.9031 & 0.8583 & 0.8035 & 0.7411 &        &        \\ \hline
(6,3) & 1.0    & 0.9991    & 0.9945 & 0.9813 & 0.9593 & 0.9284 & 0.8889 & 0.7254 & \\ \hline
(6,4) & 1.0    & 1.0    & 0.9981 & 0.9927 & 0.9834 & 0.9677 & 0.9440 & 0.8216 & \\ \hline
(7,1) & 0.9767 & 0.9366 &        &        &        &        &        &        &        \\ \hline
(7,2) & 0.9998   & 0.9790 & 0.9616 & 0.9302 & 0.8842 & 0.8264 & 0.7595 &        &        \\ \hline
(7,3) & 1.0    & 0.9998 & 0.9952 & 0.9816 & 0.9589 & 0.9271 & 0.8866 & 0.7241 &  \\ \hline
(7,4) & 1.0    & 1.0    & 0.9991 & 0.9969 & 0.9917 & 0.9818 & 0.9657 & 0.8653 & 0.7516
\end{tabular}
\caption{\textbf{4-qubit active, minimal GSM efficiencies.} The constituent BSMs are implemented using the active, QPC-encoded BSM in Ref.~\cite{lee2019fundamental}. Here, $n,m$, shown in the first column, are the considered sizes of the two repetition codes in a QPC. We assume that each photon partaking in a BSM undergoes a loss at rates listed in the first row. For every combination of $n,m$ and loss rate, we optimize the efficiency over $j$. The efficiencies far below 0.75 are omitted.}
\label{tb:Z4GHZ-ff}
\end{table*}

\section{Simulation details}
\label{app:sim}

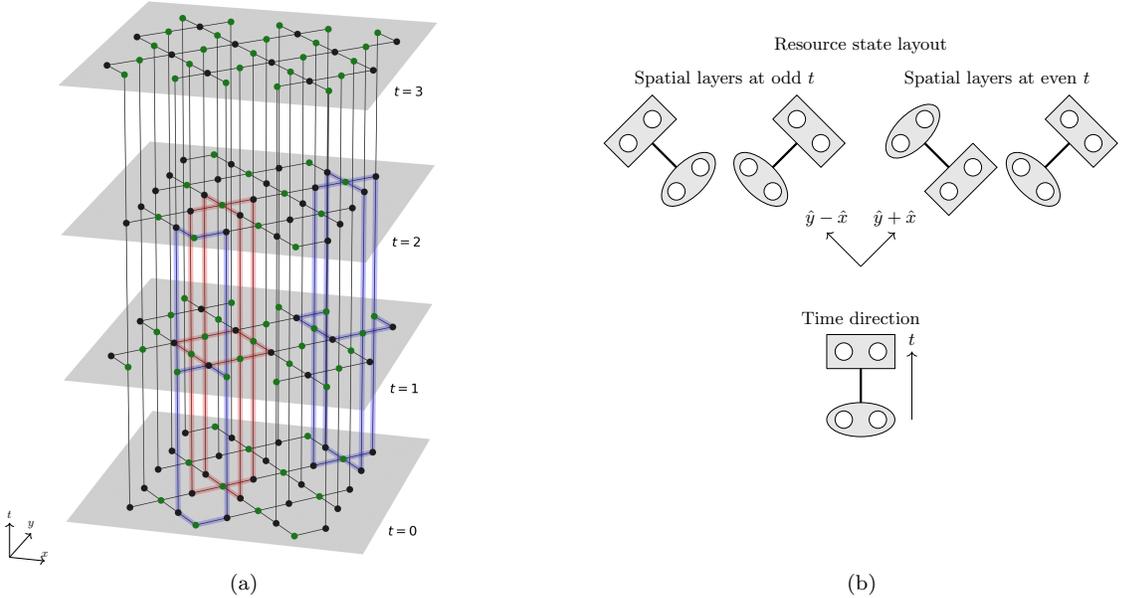
\begin{figure*}[htbp!]
    \centering
    
    \subfloat[]{\resizebox{0.6\columnwidth}{!}{
    \newcommand{\viewa}{70}
    \newcommand{\viewb}{17}
    \tdplotsetmaincoords{\viewa}{\viewb}
    
    \begin{tikzpicture}[tdplot_main_coords,thick,scale=1, every node/.style={transform shape}]
        \begin{scope}[transform canvas={scale=5.0}]
            \draw[->] (-1.25,-0,0) -- ++(1,0,0) node[anchor=south]{$x$};
            \draw[->] (-1.25,-0,0) -- ++(0,2,0) node[anchor=south]{$y$};
            \draw[->] (-1.25,-0,0) -- ++(0,0,1) node[anchor=south]{$t$};
        \end{scope}      
    \end{tikzpicture}    

    \includegraphics[scale=1]{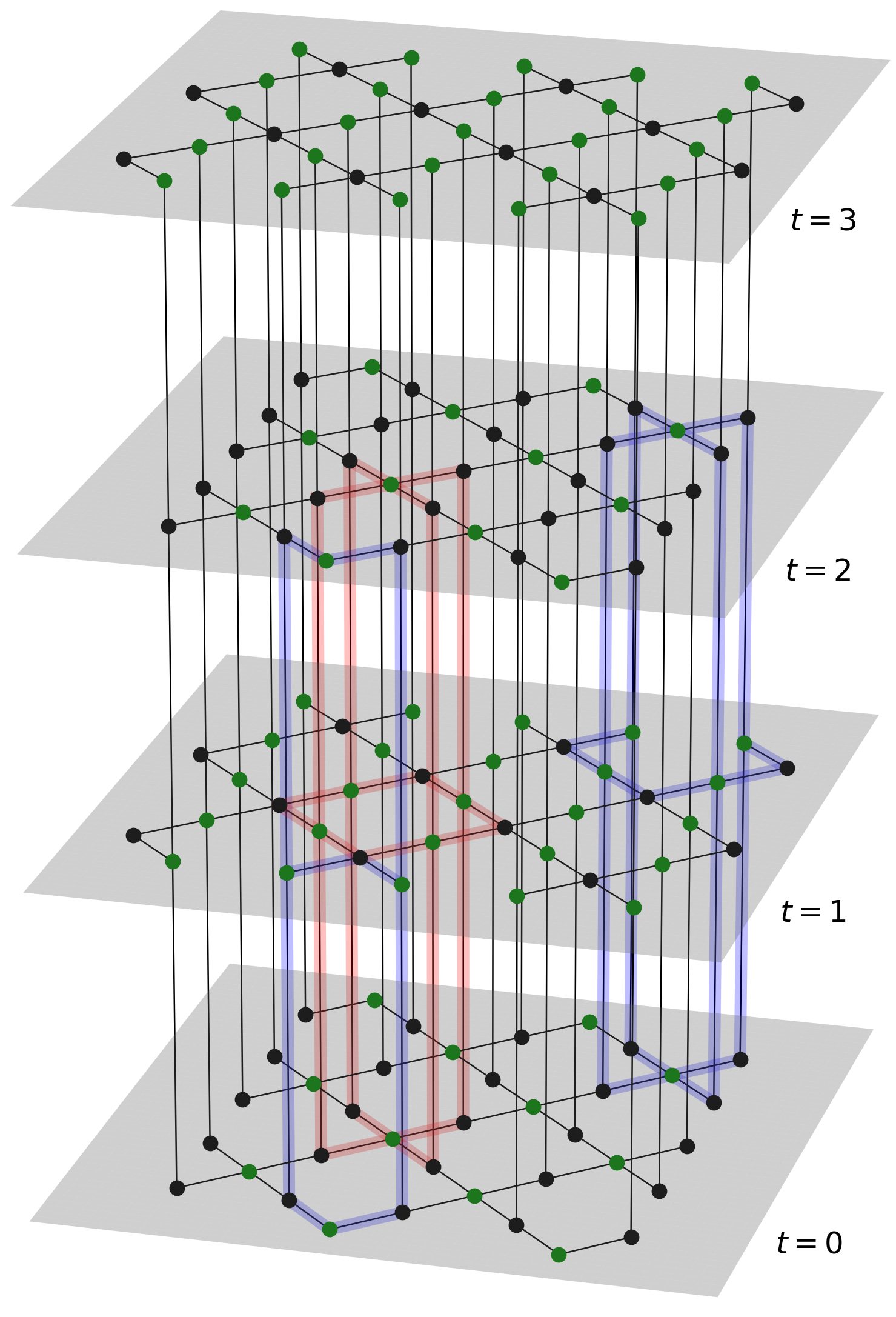}}}
    \hspace{+2cm}
    \subfloat[]{\resizebox{0.8\columnwidth}{!}{
    \begin{tikzpicture}[scale=.3,every node/.style={minimum size=0.2cm},on grid]
        \begin{scope}[yshift = 10cm]
        \node[] at (0,0) {Resource state layout};
        \end{scope}
        
        \begin{scope}[xshift = 8cm]
            \node[scale = 1] () at (0,8) {Spatial layers at even $t$};
            
            \begin{scope}[rotate = 45]
                    
                \coordinate (com) at (3,0);
            

                \fill[white!90!black] (com) ellipse ({1} and {2}); 
                \draw (com) ellipse [x radius=1,y radius=2];
                \node[fill=white,opacity=1,shape=circle,draw=black] (n1b) at ([yshift=-1cm]com) {};
                \node[fill=white,opacity=1,shape=circle,draw=black] (n1a) at ([yshift=+1cm]com) {};
                    
                \path[shape=coordinate]
                    ([xshift=+3cm,yshift=+2cm]com) coordinate(b1) ([xshift=+5cm,yshift=+2cm]com) coordinate(b2)
                    ([xshift=+5cm,yshift=-2cm]com) coordinate(b3) ([xshift=+3cm,yshift=-2cm]com) coordinate(b4);
                \filldraw[fill=white!90!black] (b1) -- (b2) -- (b3) -- (b4) -- (b1);
                \node[fill=white,opacity=1,shape=circle,draw=black] (n2b) at ([xshift=+4cm,yshift=+1cm]com) {};
                \node[fill=white,opacity=1,shape=circle,draw=black] (n2a) at ([xshift=+4cm,yshift=-1cm]com) {};
            
                \begin{pgfonlayer}{background}
                    \path [-,line width=0.04cm,black] ([xshift=+4cm]com) edge node {} (com);
                \end{pgfonlayer}
         
            \end{scope}
        
            \begin{scope}[rotate = -45]
                    
                \coordinate (com) at (-7,0);
            

                \fill[white!90!black] (com) ellipse ({1} and {2}); 
                \draw (com) ellipse [x radius=1,y radius=2];
                \node[fill=white,opacity=1,shape=circle,draw=black] (n1b) at ([yshift=-1cm]com) {};
                \node[fill=white,opacity=1,shape=circle,draw=black] (n1a) at ([yshift=+1cm]com) {};
                    
                \path[shape=coordinate]
                    ([xshift=+3cm,yshift=+2cm]com) coordinate(b1) ([xshift=+5cm,yshift=+2cm]com) coordinate(b2)
                    ([xshift=+5cm,yshift=-2cm]com) coordinate(b3) ([xshift=+3cm,yshift=-2cm]com) coordinate(b4);
                \filldraw[fill=white!90!black] (b1) -- (b2) -- (b3) -- (b4) -- (b1);
                \node[fill=white,opacity=1,shape=circle,draw=black] (n2b) at ([xshift=+4cm,yshift=+1cm]com) {};
                \node[fill=white,opacity=1,shape=circle,draw=black] (n2a) at ([xshift=+4cm,yshift=-1cm]com) {};
            
                \begin{pgfonlayer}{background}
                    \path [-,line width=0.04cm,black] ([xshift=+4cm]com) edge node {} (com);
                \end{pgfonlayer}

            \end{scope}
    
            \node[] () at (0,-2) {};
        \end{scope}
    
        \begin{scope}[xshift = -8cm]
            \node[scale = 1] () at (0,8) {Spatial layers at odd $t$};
            
            \begin{scope}[rotate = 45]
                    
                \coordinate (com) at (3,0);
            

                \fill[white!90!black] (com) ellipse ({1} and {2}); 
                \draw (com) ellipse [x radius=1,y radius=2];
                \node[fill=white,opacity=1,shape=circle,draw=black] (n1b) at ([yshift=-1cm]com) {};
                \node[fill=white,opacity=1,shape=circle,draw=black] (n1a) at ([yshift=+1cm]com) {};
                    
                \path[shape=coordinate]
                    ([xshift=+3cm,yshift=+2cm]com) coordinate(b1) ([xshift=+5cm,yshift=+2cm]com) coordinate(b2)
                    ([xshift=+5cm,yshift=-2cm]com) coordinate(b3) ([xshift=+3cm,yshift=-2cm]com) coordinate(b4);
                \filldraw[fill=white!90!black] (b1) -- (b2) -- (b3) -- (b4) -- (b1);
                \node[fill=white,opacity=1,shape=circle,draw=black] (n2b) at ([xshift=+4cm,yshift=+1cm]com) {};
                \node[fill=white,opacity=1,shape=circle,draw=black] (n2a) at ([xshift=+4cm,yshift=-1cm]com) {};
            
                \begin{pgfonlayer}{background}
                    \path [-,line width=0.04cm,black] ([xshift=+4cm]com) edge node {} (com);
                \end{pgfonlayer}
         
            \end{scope}
        
            \begin{scope}[rotate = 135]
                    
                \coordinate (com) at (3,0);
            

                \fill[white!90!black] (com) ellipse ({1} and {2}); 
                \draw (com) ellipse [x radius=1,y radius=2];
                \node[fill=white,opacity=1,shape=circle,draw=black] (n1b) at ([yshift=-1cm]com) {};
                \node[fill=white,opacity=1,shape=circle,draw=black] (n1a) at ([yshift=+1cm]com) {};
                    
                \path[shape=coordinate]
                    ([xshift=+3cm,yshift=+2cm]com) coordinate(b1) ([xshift=+5cm,yshift=+2cm]com) coordinate(b2)
                    ([xshift=+5cm,yshift=-2cm]com) coordinate(b3) ([xshift=+3cm,yshift=-2cm]com) coordinate(b4);
                \filldraw[fill=white!90!black] (b1) -- (b2) -- (b3) -- (b4) -- (b1);
                \node[fill=white,opacity=1,shape=circle,draw=black] (n2b) at ([xshift=+4cm,yshift=+1cm]com) {};
                \node[fill=white,opacity=1,shape=circle,draw=black] (n2a) at ([xshift=+4cm,yshift=-1cm]com) {};
            
                \begin{pgfonlayer}{background}
                    \path [-,line width=0.04cm,black] ([xshift=+4cm]com) edge node {} (com);
                \end{pgfonlayer}
         
            \end{scope}
    
            \node[] () at (0,-2) {};
        \end{scope}

        \begin{scope}[yshift = -3cm]
            \draw[scale = 1,->] (0,0) -- ++(2,2) node[anchor=south]{$\hat{y}+\hat{x}$};
            \draw[scale = 1,->] (0,0) -- ++(-2,2) node[anchor=south]{$\hat{y}-\hat{x}$};
        \end{scope}
    
        \begin{scope}[yshift = -12cm]
    
            \node[scale = 1] () at (0,8) {};
            
            \node[scale = 1] () at (0,6) {Time direction};
            
            \begin{scope}[rotate = 90]
                    
                \coordinate (com) at (0,0);
            

                \fill[white!90!black] (com) ellipse ({1} and {2}); 
                \draw (com) ellipse [x radius=1,y radius=2];
                \node[fill=white,opacity=1,shape=circle,draw=black] (n1b) at ([yshift=-1cm]com) {};
                \node[fill=white,opacity=1,shape=circle,draw=black] (n1a) at ([yshift=+1cm]com) {};
                    
                \path[shape=coordinate]
                    ([xshift=+3cm,yshift=+2cm]com) coordinate(b1) ([xshift=+5cm,yshift=+2cm]com) coordinate(b2)
                    ([xshift=+5cm,yshift=-2cm]com) coordinate(b3) ([xshift=+3cm,yshift=-2cm]com) coordinate(b4);
                \filldraw[fill=white!90!black] (b1) -- (b2) -- (b3) -- (b4) -- (b1);
                \node[fill=white,opacity=1,shape=circle,draw=black] (n2b) at ([xshift=+4cm,yshift=+1cm]com) {};
                \node[fill=white,opacity=1,shape=circle,draw=black] (n2a) at ([xshift=+4cm,yshift=-1cm]com) {};
            
                \begin{pgfonlayer}{background}
                    \path [-,line width=0.04cm,black] ([xshift=+4cm]com) edge node {} (com);
                \end{pgfonlayer}
         
            \end{scope}
        
            \draw[scale = 2,->] (1.5,0) -- ++(0,2) node[anchor=south]{$t$};

            \node[] at (0,-8) {};
            
        \end{scope}

    \end{tikzpicture}}
    \vspace{10cm}}
    \caption{(a) An example of a graph defined by a foliated, rotated distance-5 surface code (only four time slices are shown for visualization.). 
    (b) The resource-state (encoded two-qubit graph state) lay-out with respect to the graph in (a). (See Fig.~\ref{fig:unit_cell} for the definition of the resource states.) As a result, the number of (encoded) qubits at each vertex is given by the degree of the vertex.
    The subgraph highlighted in red depicts a RHG unit cell, whereas the subgraphs highlighted in blue are examples of partial unit cells at the boundaries.
    For each unit cell, complete or partial, the associated parity check operator is a product of $\prod_{i=1}^m X_i$-type GSM outcome at the green vertices, where $m$ is the vertex-degree, and $ZZ$-type GSM outcome at the black vertices.}
    \label{fig:fnet}
\end{figure*}

Here we provide details on the numerical simulations used to estimate the photon-loss thresholds. In the simulations, we consider fusion networks based on a graph $\mathcal{G}$ defined by a foliated, rotated surface code with boundaries, depicted in Fig.~\ref{fig:fnet}. While the bulk of $\mathcal{G}$ can be tiled by cubic unit cells, at the boundaries, parts of the unit cells will be ``cut off". Crucially, one can still define parity check operators for these partial unit cells, as we show in Fig.~\ref{fig:fnet}. 

\begin{figure*}
\centering
\subfloat[before edge (erasure) contraction]{
	\label{subfig:before_merging}
\includegraphics[width = \columnwidth]{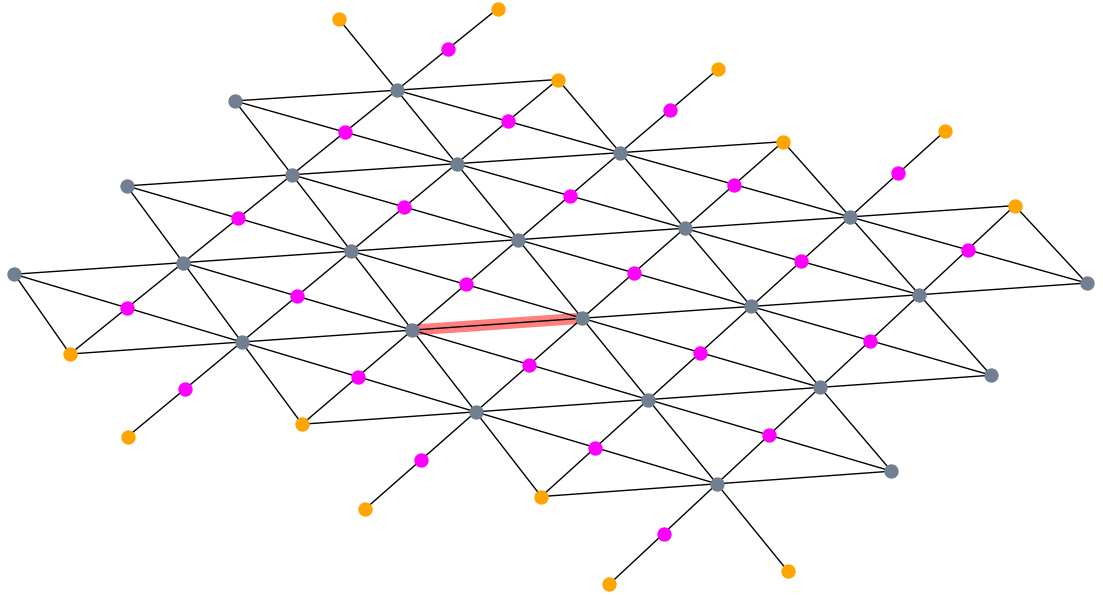}}
\subfloat[after edge contraction]{
	\label{subfig:after_merging}
\includegraphics[width = \columnwidth]{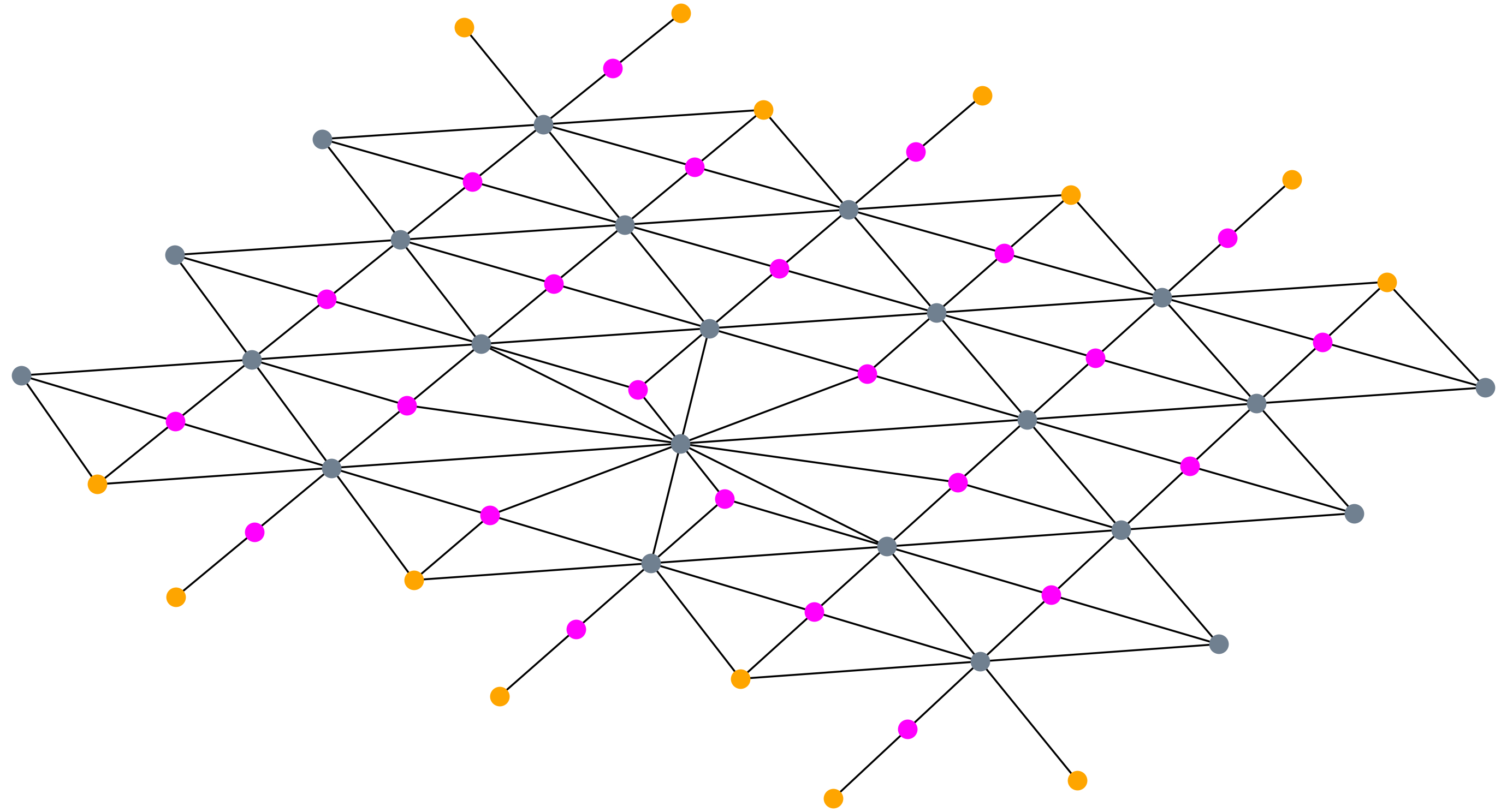}}
\caption{Left: The highlighted red edge, which represents an $XXXX$-type GSM outcome in a fusion network, is erased (left). Right: The gray vertices, which represent cubic parity check operators, at the endpoints of the red edge are merged to form a check operator, which is independent of the erased outcome, on a spatial layer of the cyclic primal syndrome graph (time-like edges are ignored for visualisation). The purple vertices are the parity check operators which are the product of all $ZZ$-type outcomes of a GSM at an edge of a cubic unit cell (see Sec.~\ref{sec:QEC}); the orange vertices are the boundary vertices.}
\label{fig:contracting_nodes_cyclic_syn_graph}
\end{figure*}

\begin{figure*}
\centering
\subfloat[]{\centering \resizebox{0.25\textwidth}{!}{\includegraphics[]{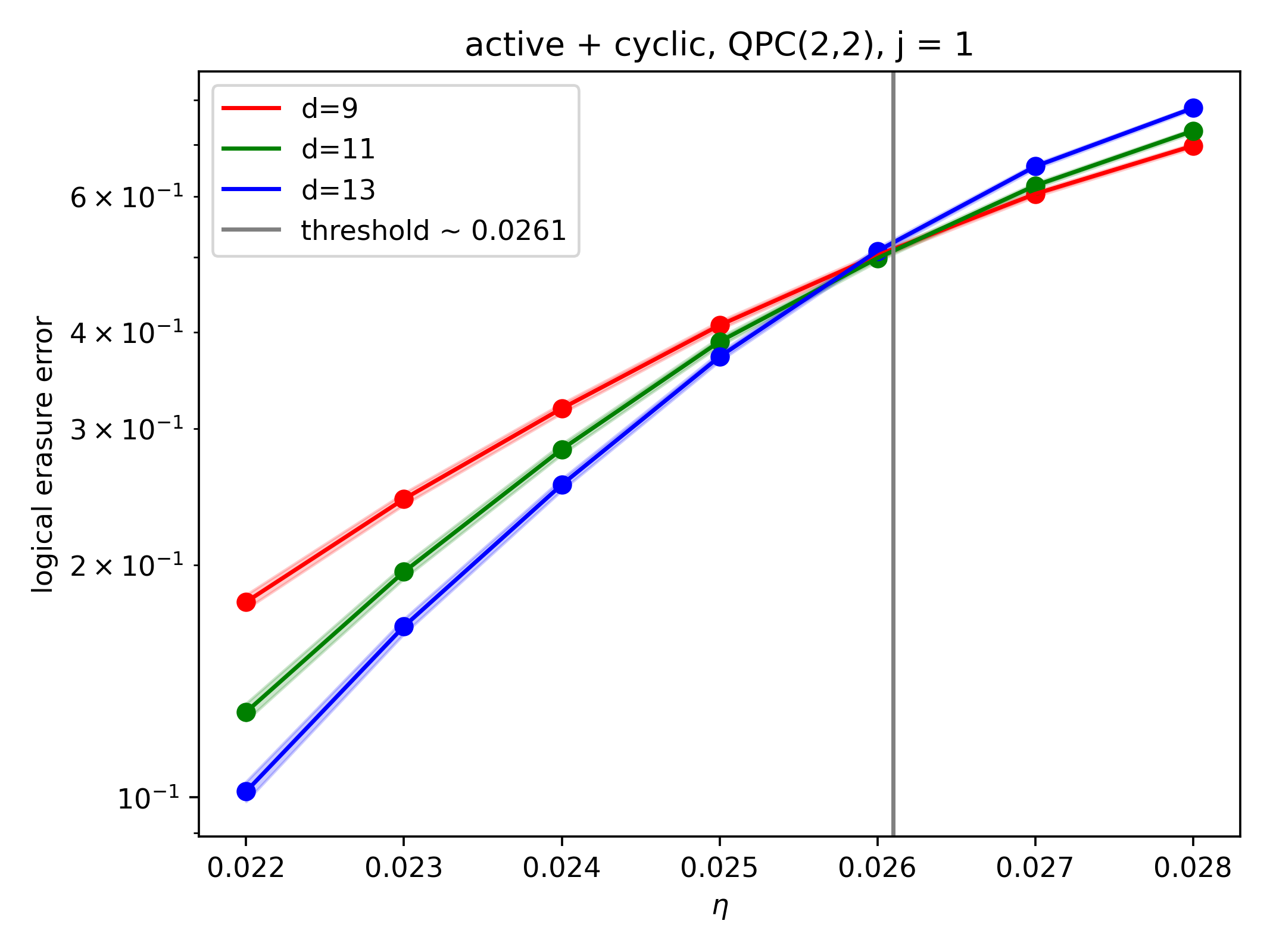}}}
\subfloat[]{\centering \resizebox{0.25\textwidth}{!}{\includegraphics[]{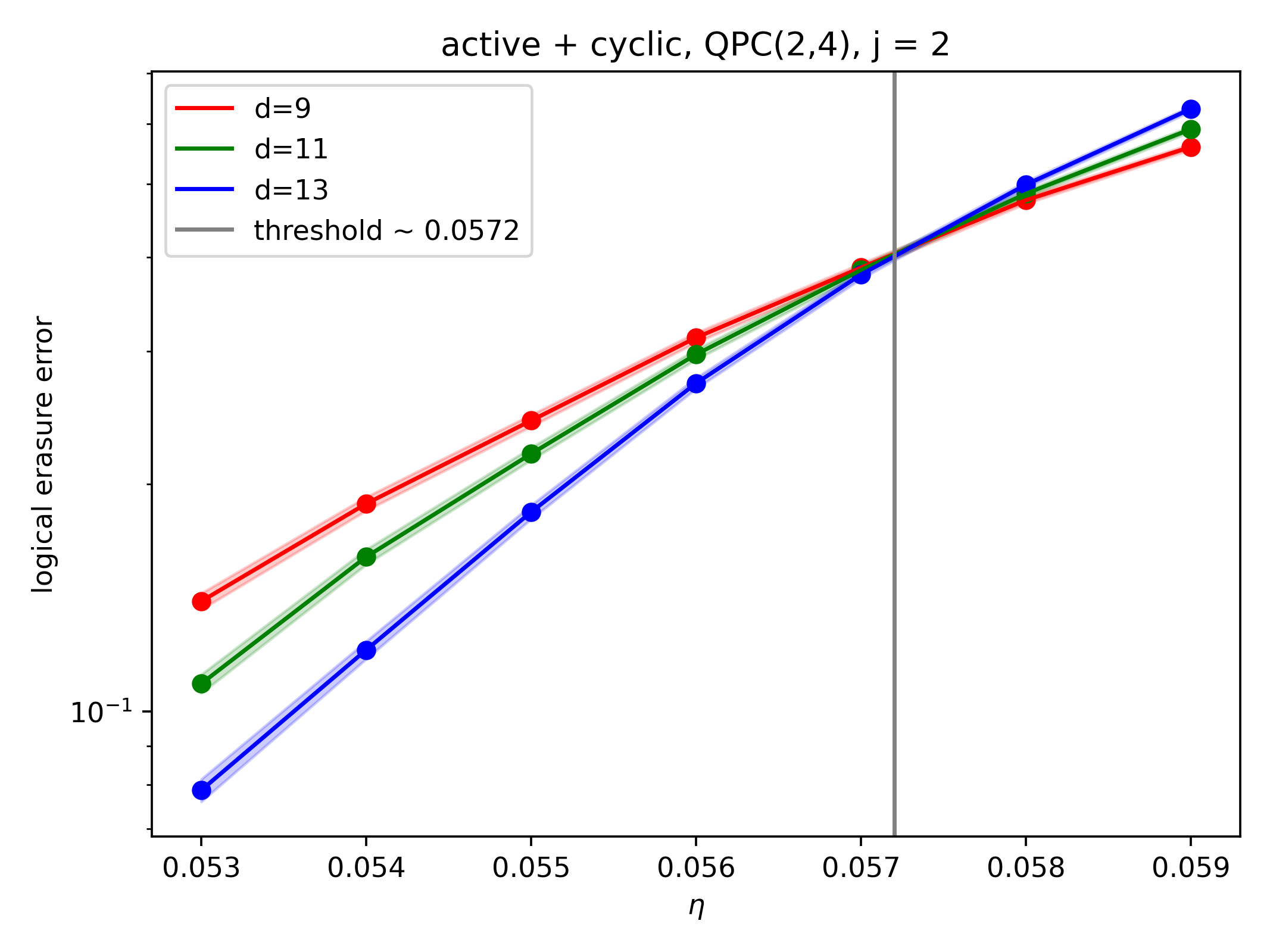}}}
\subfloat[]{\centering \resizebox{0.25\textwidth}{!}{\includegraphics[]{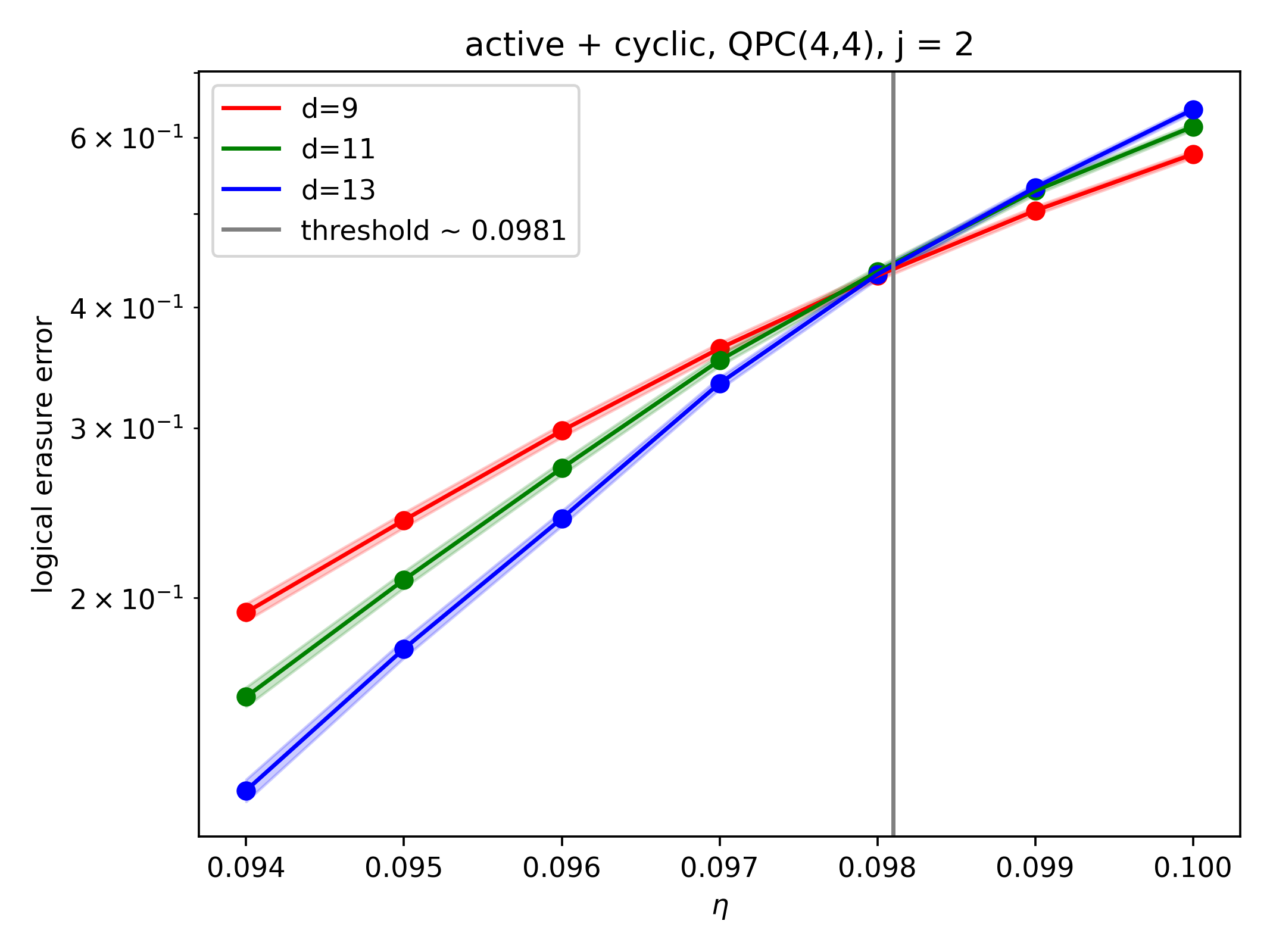}}}
\subfloat[]{\centering \resizebox{0.25\textwidth}{!}{\includegraphics[]{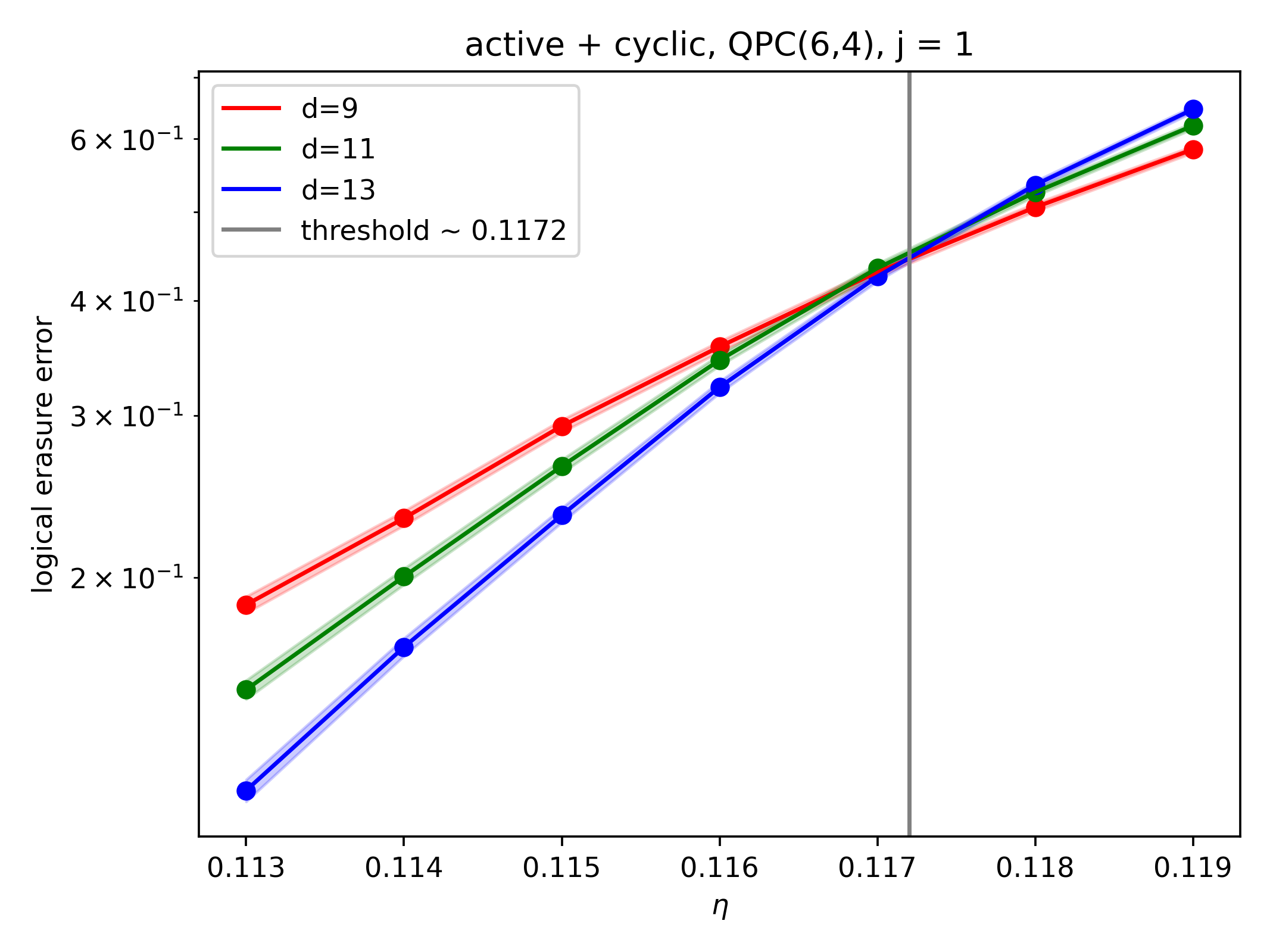}}}

\vspace{-0.1cm}
\subfloat[]{\centering \resizebox{0.25\textwidth}{!}{\includegraphics[]{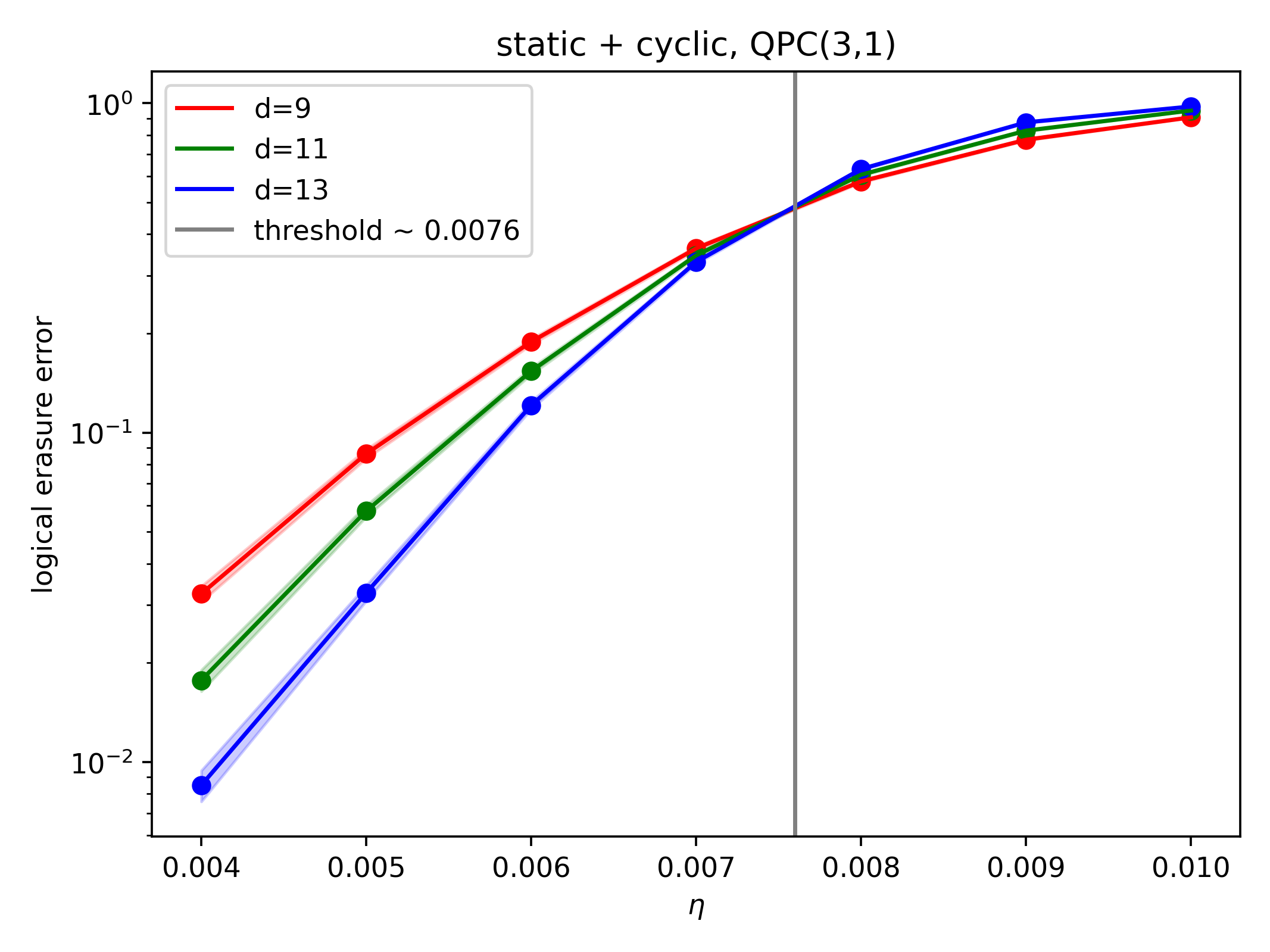}}}
\subfloat[]{\centering \resizebox{0.25\textwidth}{!}{\includegraphics[]{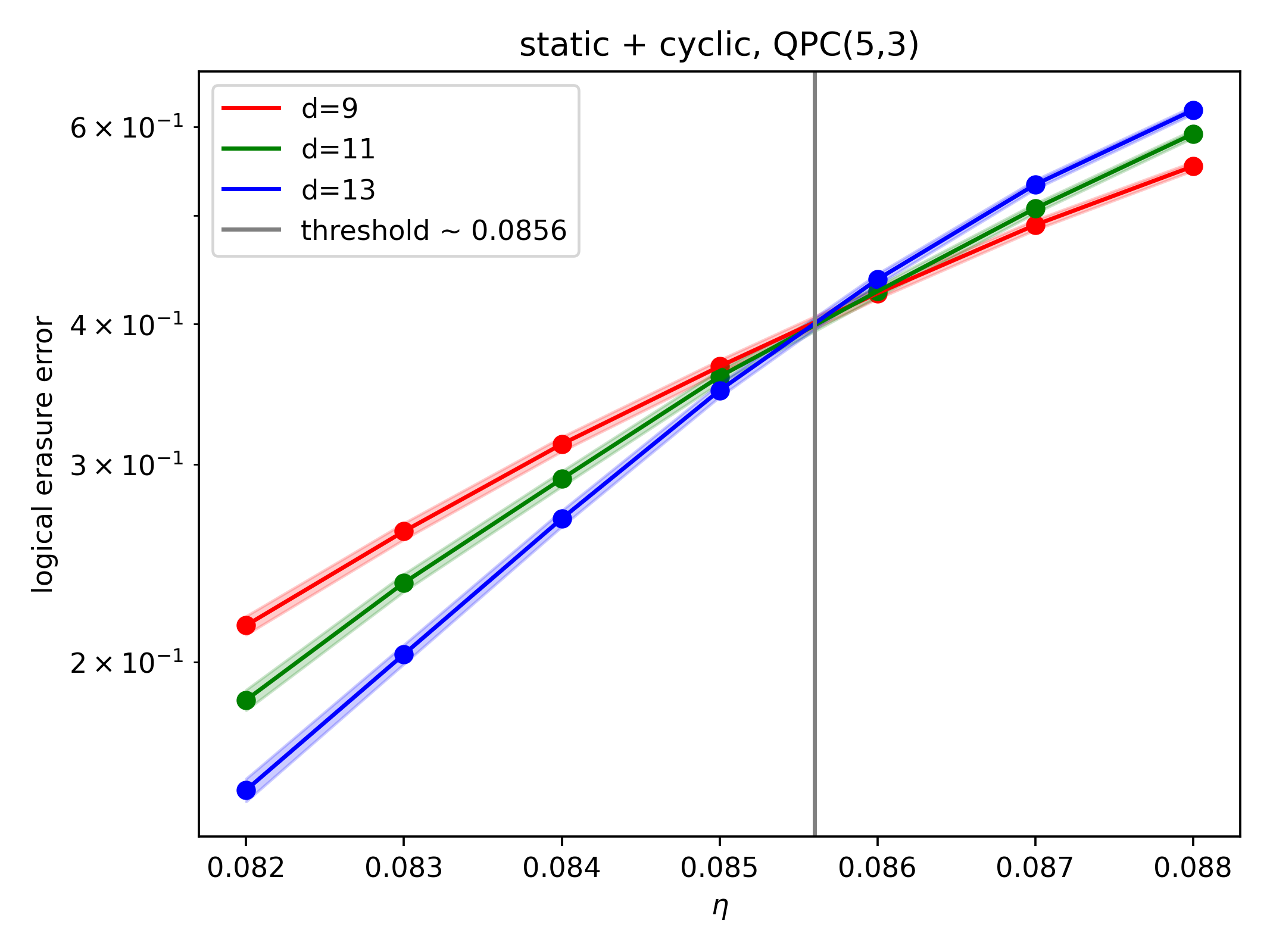}}}
\subfloat[]{\centering \resizebox{0.25\textwidth}{!}{\includegraphics[]{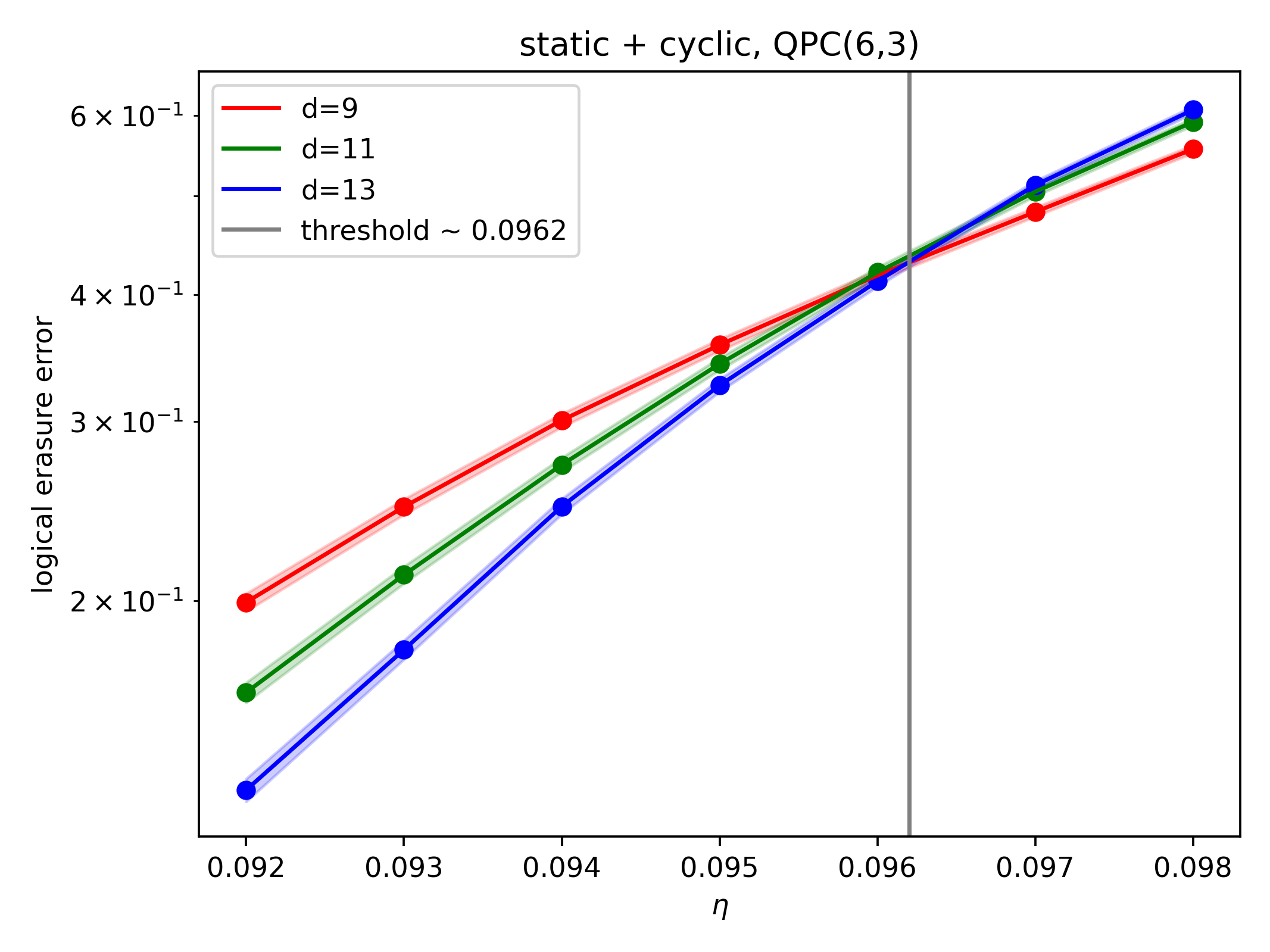}}}
\subfloat[]{\centering \resizebox{0.25\textwidth}{!}{\includegraphics[]{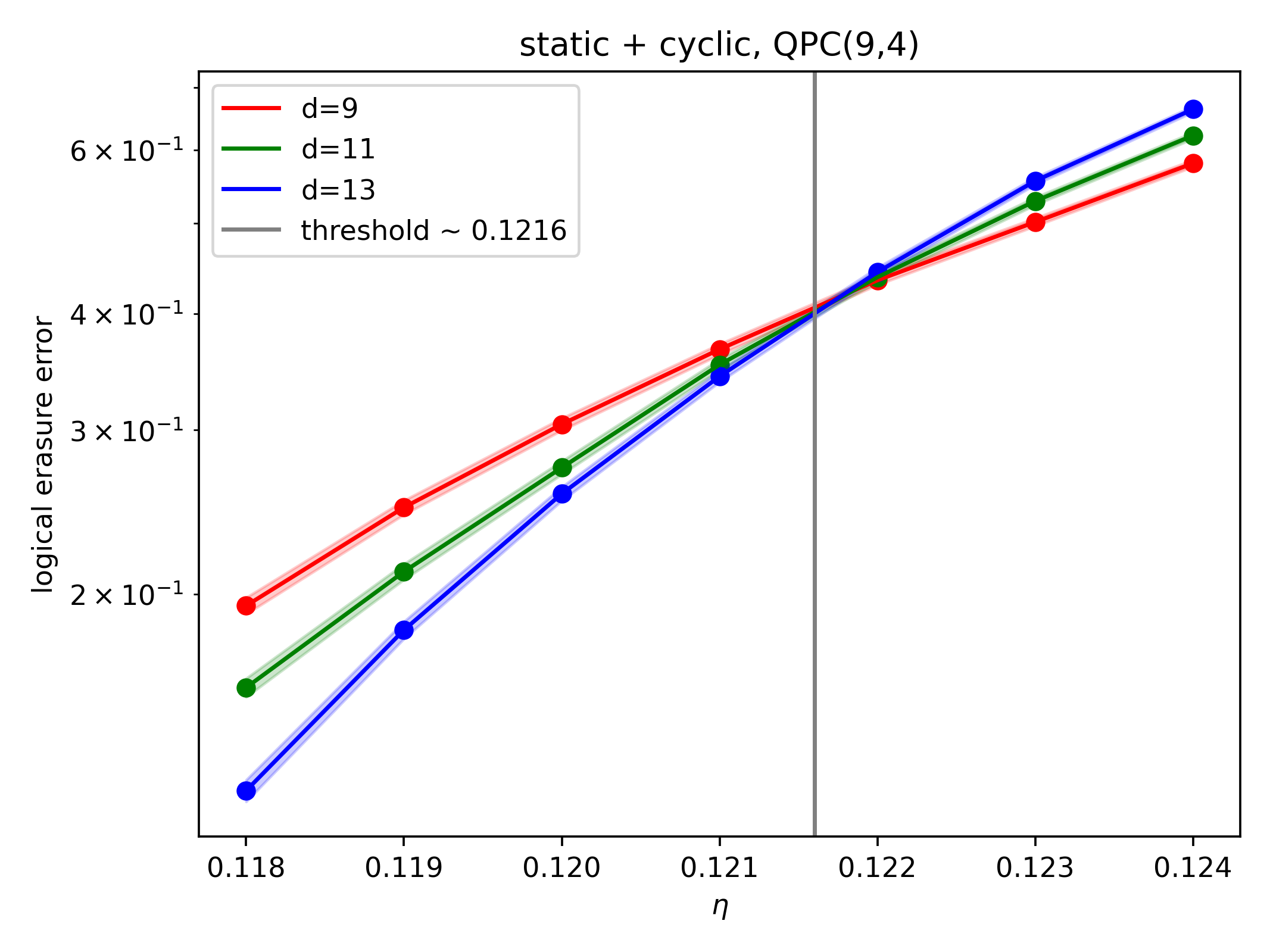}}}

\vspace{-0.1cm}
\subfloat[]{\centering \resizebox{0.25\textwidth}{!}{\includegraphics[]{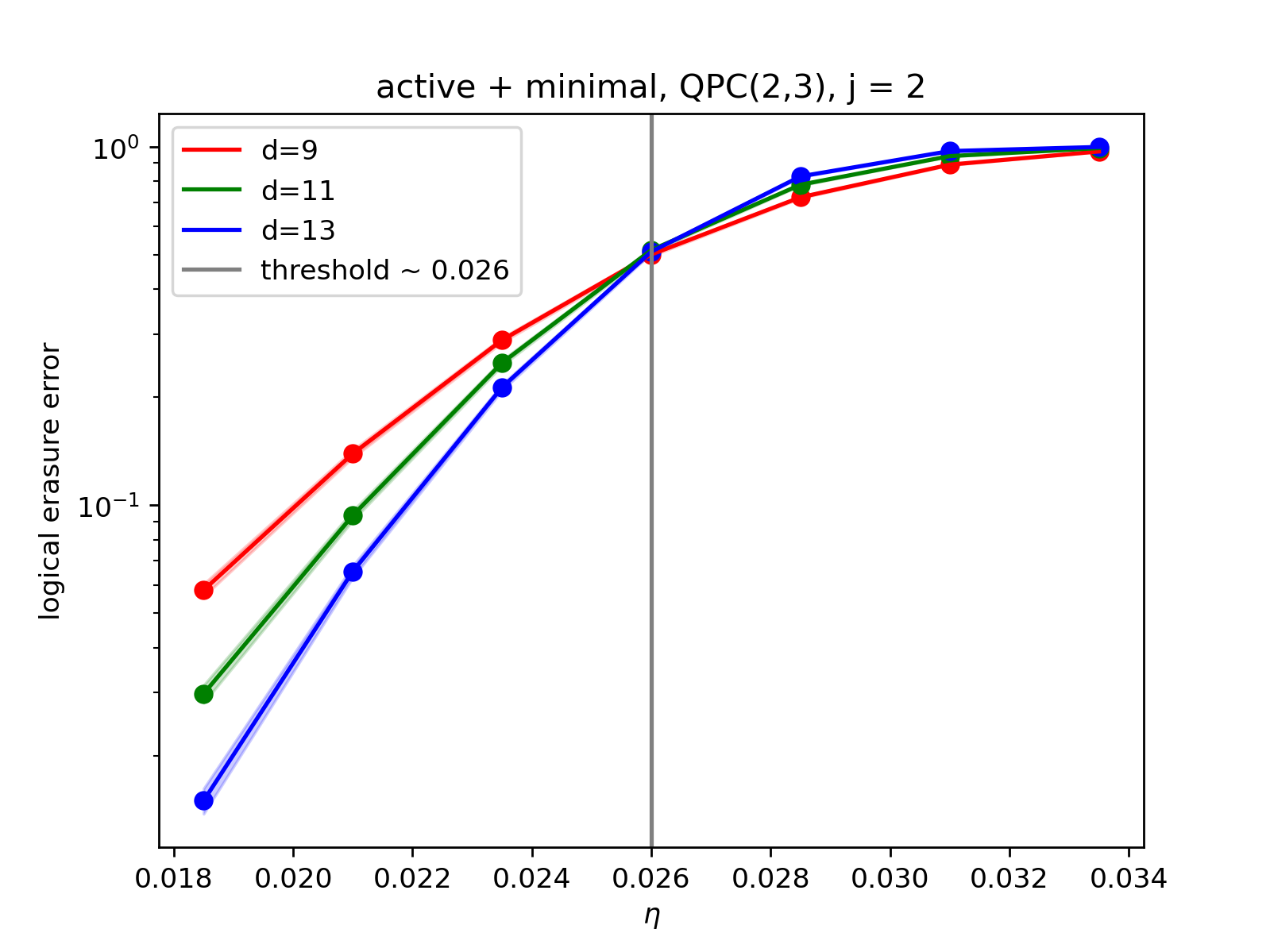}}}
\subfloat[]{\centering \resizebox{0.25\textwidth}{!}{\includegraphics[]{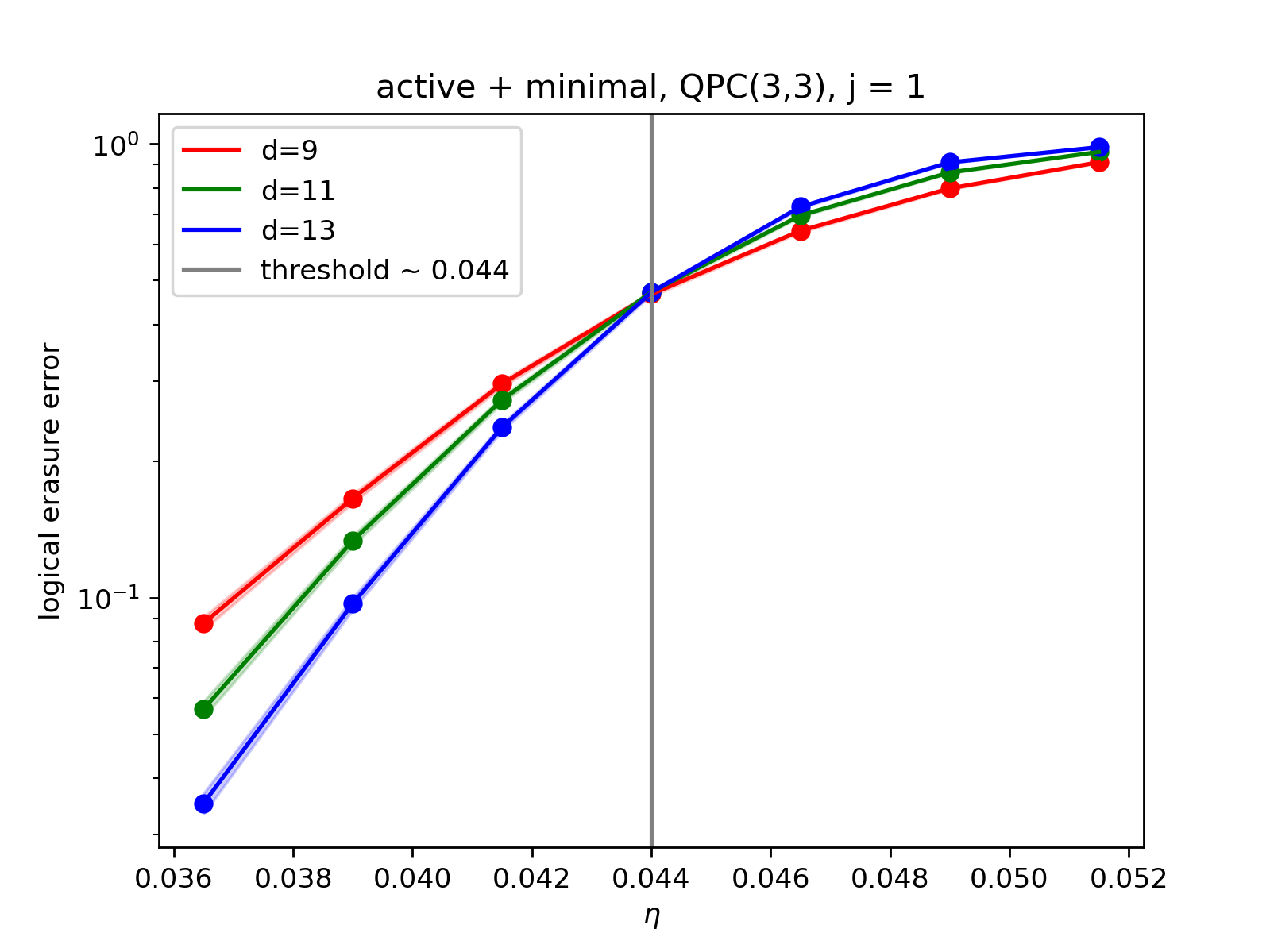}}}
\subfloat[]{\centering \resizebox{0.25\textwidth}{!}{\includegraphics[]{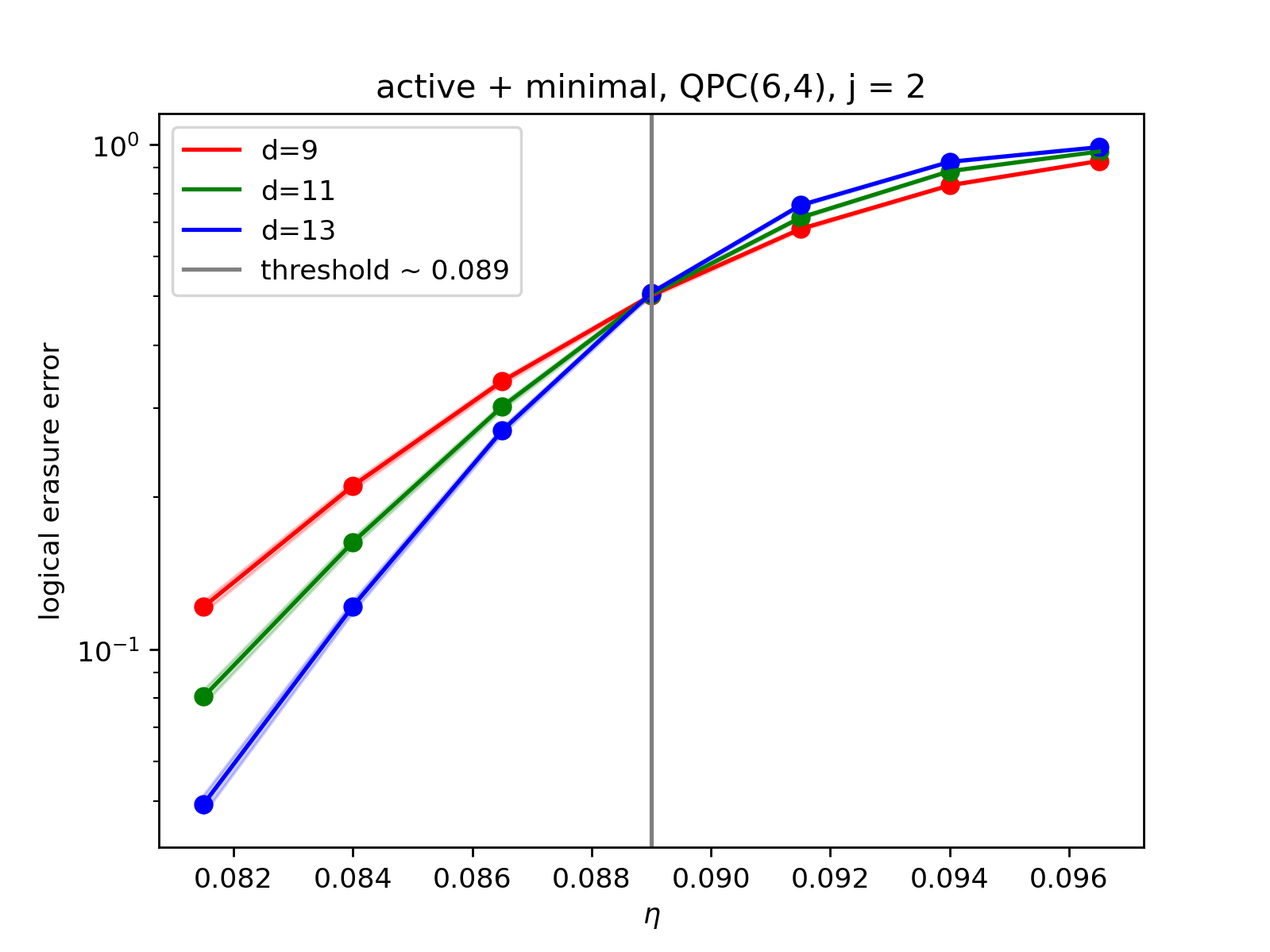}}}
\subfloat[]{\centering \resizebox{0.25\textwidth}{!}{\includegraphics[]{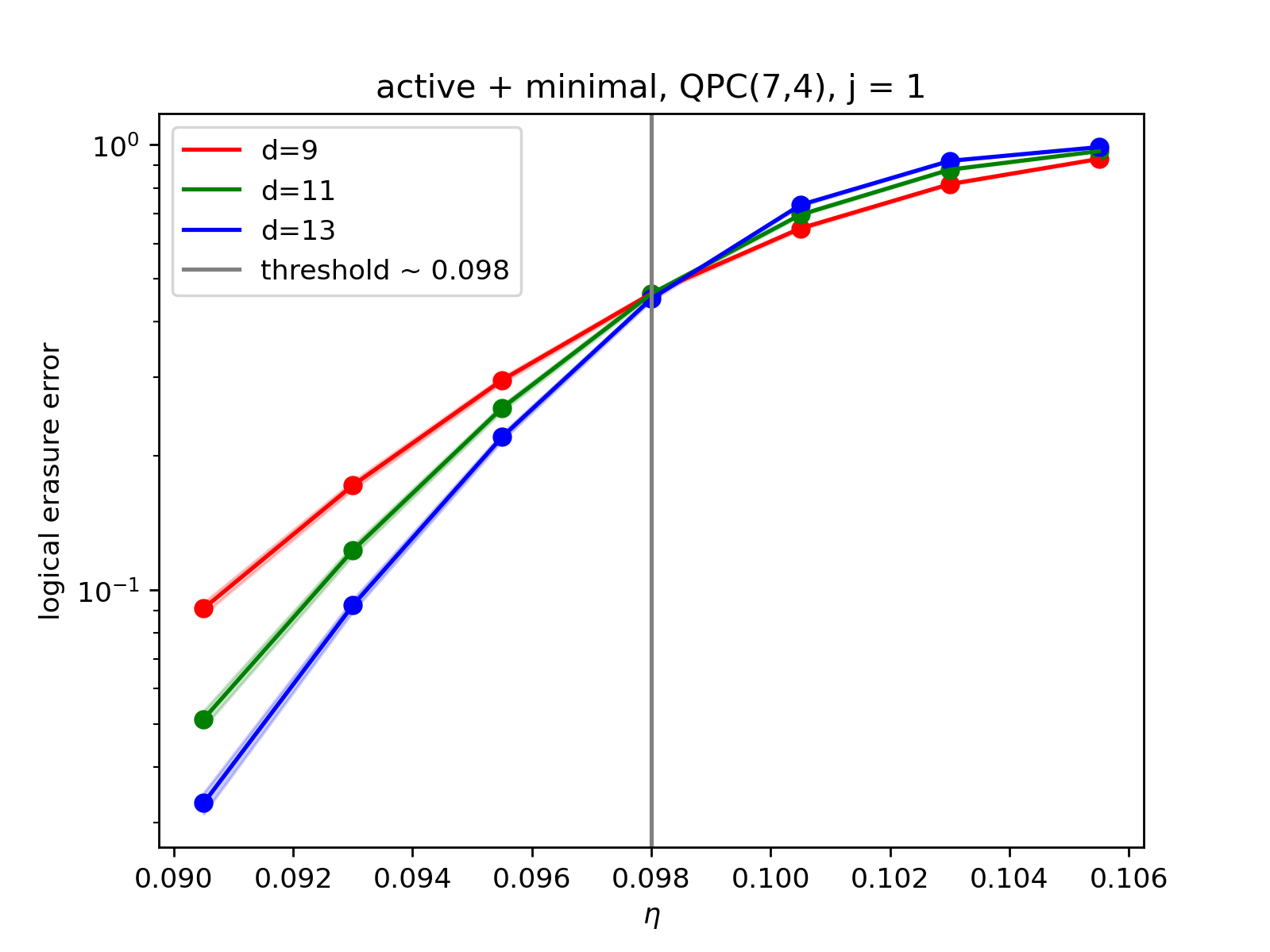}}}

\vspace{-0.1cm}
\subfloat[]{\centering \resizebox{0.25\textwidth}{!}{\includegraphics[]{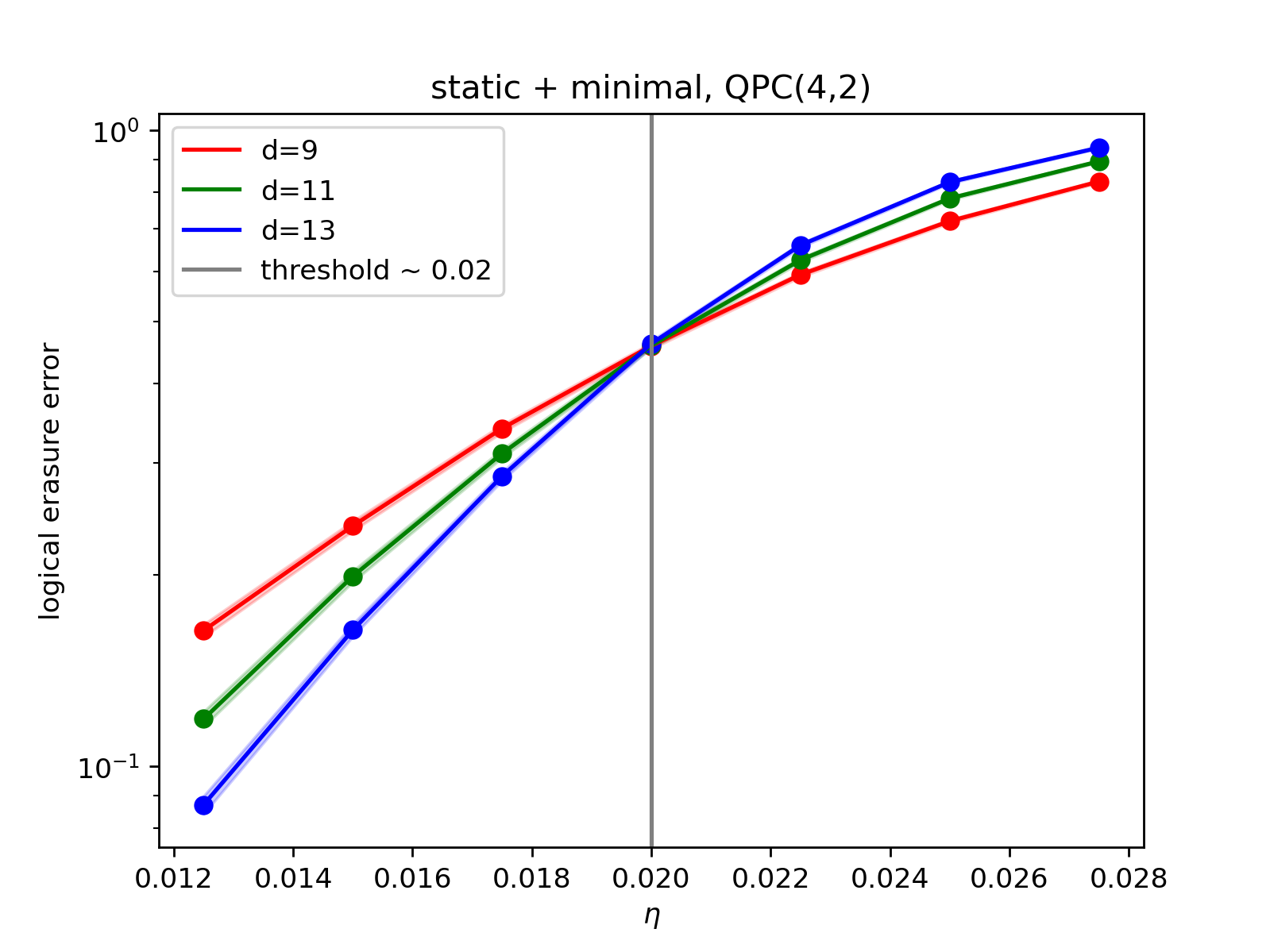}}}
\subfloat[]{\centering \resizebox{0.25\textwidth}{!}{\includegraphics[]{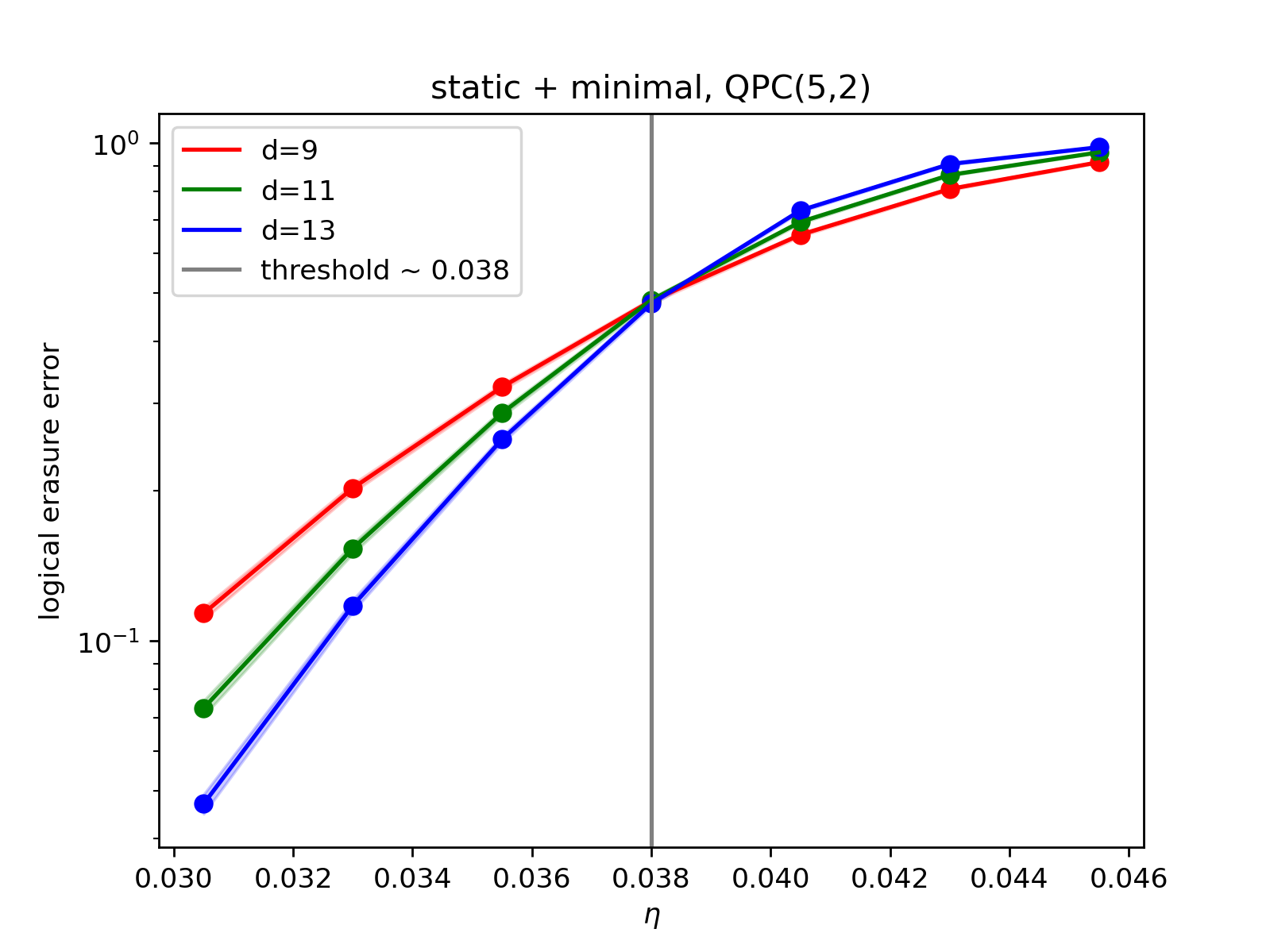}}}
\subfloat[]{\centering \resizebox{0.25\textwidth}{!}{\includegraphics[]{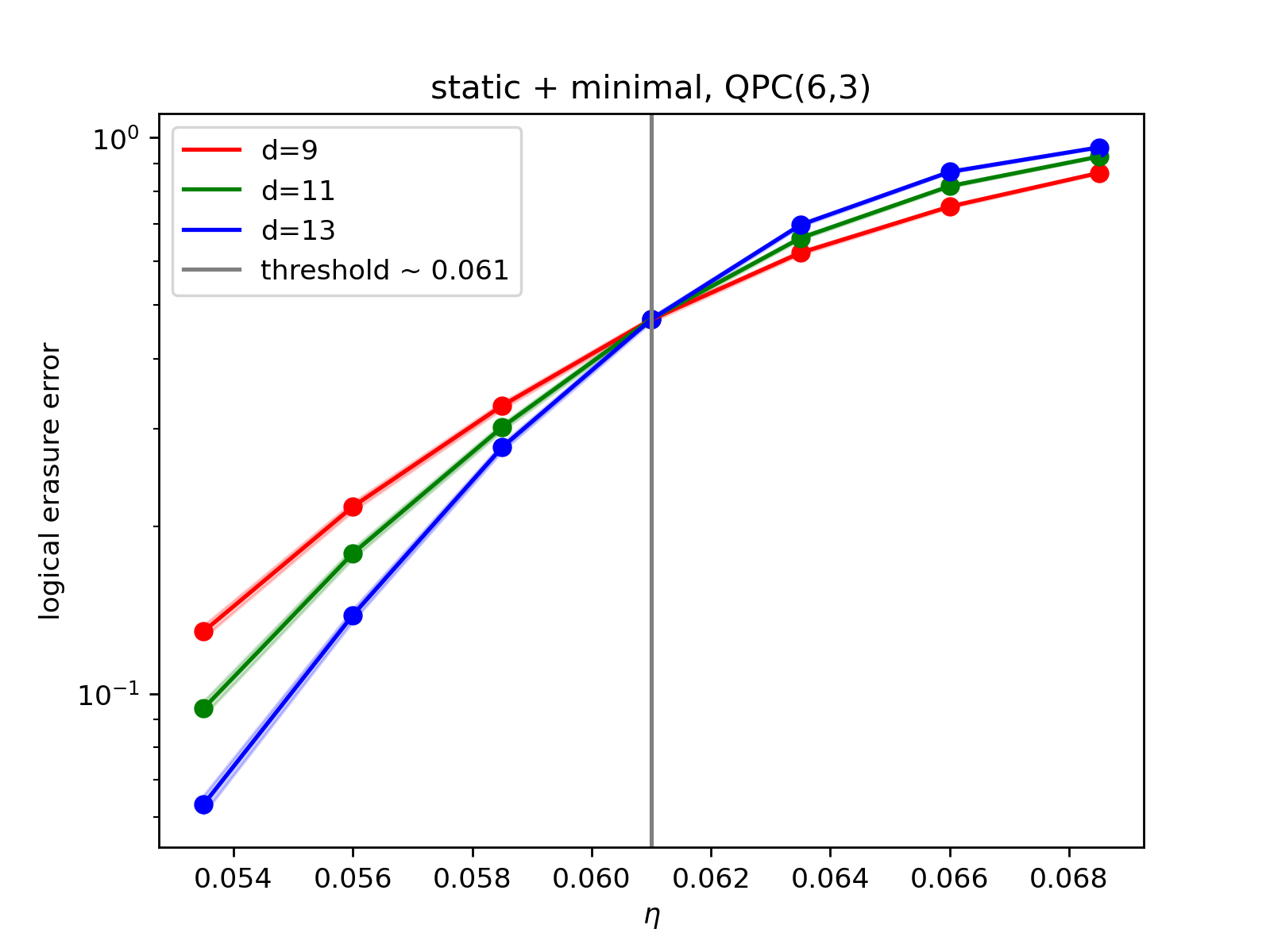}}}
\subfloat[]{\centering \resizebox{0.25\textwidth}{!}{\includegraphics[]{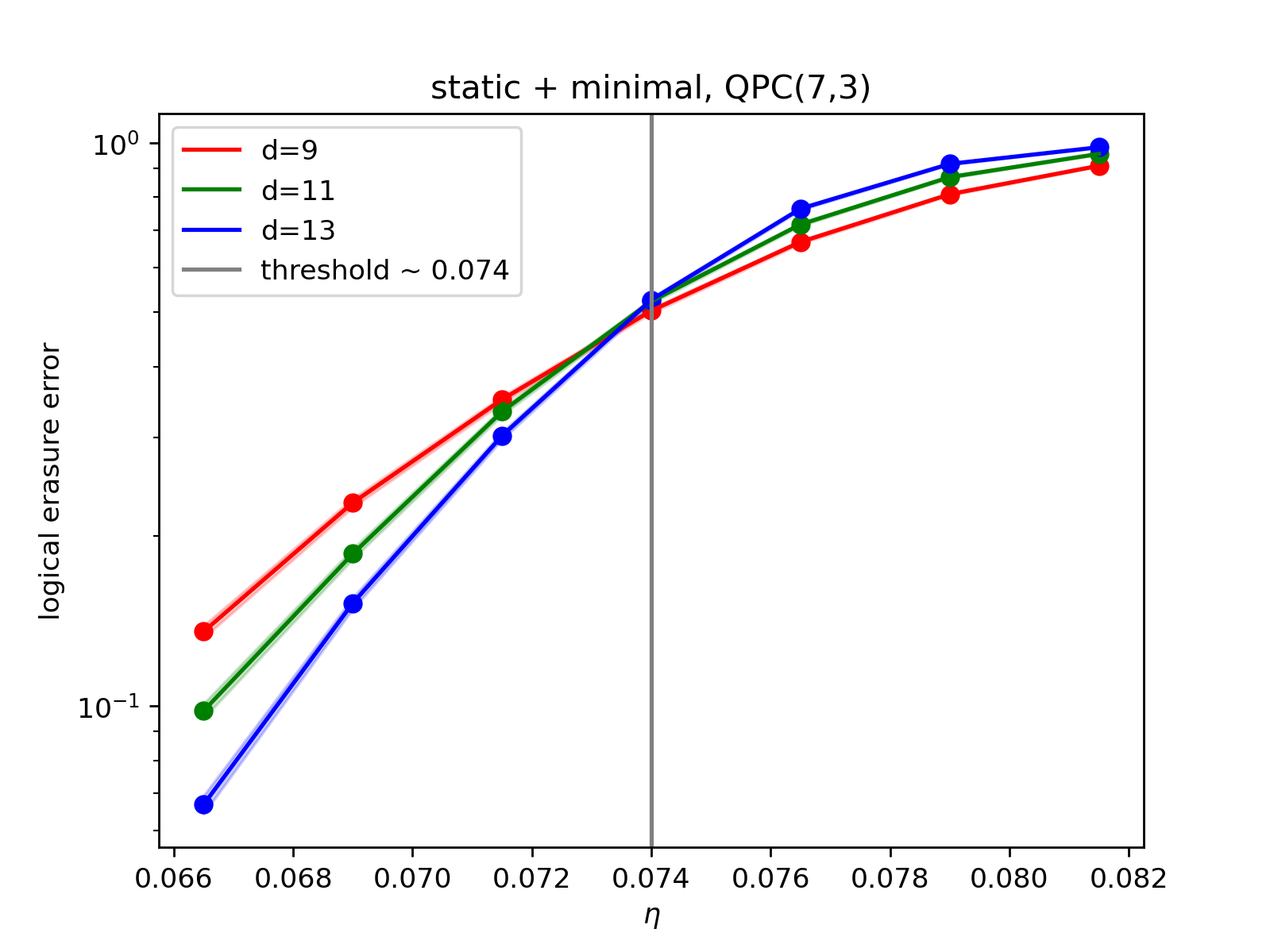}}}

\label{fig:threshold_ex}
\caption{Examples of single-photon loss threshold estimates. (a)-(d) correspond to thresholds for the cyclic architecture constructed using active entangling measurements.(e)-(h) correspond to thresholds for the cyclic architecture constructed using static entangling measurements. (i)-(l) correspond to thresholds for the minimal architecture constructed using active entangling measurements. (m)-(p) correspond to thresholds for the minimal architecture constructed using static entangling measurements.}
\end{figure*}

We perform Monte-Carlo simulations to estimate the thresholds under the linear-optical error model, where the erasure probabilities of different measurement outcomes are determined by the following parameters (see App.~\ref{app:ghz_erasure} for more details):
\begin{itemize}
    \item the choice of architecture: minimal or cyclic,
    \item the choice of encoded BSM: static~\cite{ewert2016ultrafast} or active~\cite{lee2019fundamental} protocol (see App.~\ref{app:QPC_BSM} for details.),
    \item single-photon loss rate $\eta$,
    \item QPC code parameters $n,m$,
    \item feed-forward parameter $j$ if the active encoded BSM is chosen,
    \item the choice of QPC definition: Eqn.~\eqref{eq:qpc_had} or Eqn.~\eqref{eq:qpc_shor}.
\end{itemize}
Note that while all bulk vertices have degree four, some boundary vertices have degree two or three. Therefore, a fusion network comprises 2, 3 and 4-qubit GSMs, leading to four possible types of GSM outcomes: $XX$, $XXX$, $XXXX$, and $ZZ$, each of which is associated with a different erasure probability given a combination of parameters. For each combination of parameters, we simulate fusion networks of dimensions $d\times d \times 2d+1$, where $d$ is the distance of the underlying surface code, for various $d \in \{9,11,13\}$. In each Monte-Carlo sample, we assign erasure errors on the primal and dual syndrome graphs according to the probabilities discussed above, and compute the resulting syndrome graphs. We illustrate in Fig.~\ref{fig:contracting_nodes_cyclic_syn_graph} how an erasure error modifies a syndrome graph. We perform decoding on the primal and dual syndrome graphs separately by inspecting if a chain of erasure errors has connected two opposite boundaries, i.e., the errors have percolated; if it is true in either the primal or dual syndrome graph, a logical error has occurred~\cite{fowler2009topological,whiteside2014upper}. Alternatively, the linear-time erasure decoder~\cite{Delfosse2020linear} can be used. We assume no erasures occur in the first and last time-slices, as in Ref.~\cite{auger2018fault,omkar2022all}. For each combination of parameters, we perform and decode $10^4$ samples to compute the logical error rate. We sweep the values of $\eta$ and construct curves of $\eta$ versus logical error rate for $d \in \{9,11,13\}$ (see Fig.~\ref{fig:threshold_ex} for examples.). We then estimate the single-photon loss threshold, which constitutes a point in Fig.~\ref{fig:threshold}, as the crossing of the curves.

In table~\ref{tb:thresholds}, we list the thresholds plotted in Fig.~\ref{fig:threshold} and their corresponding parameters $n,m$ and $j$ only when the adaptive BSMs are used. Furthermore, we obtained higher thresholds for the cyclic architecture if we use the QPC definition in Eqn.~\eqref{eq:qpc_shor}, whereas the QPC definition in Eqn.~\eqref{eq:qpc_had} leads to higher thresholds for the minimal architecture. Therefore, we adopt the QPC definition in Eqn.~\eqref{eq:qpc_shor} and~\eqref{eq:qpc_had} for the curves corresponding to the cyclic and minimal architecture, respectively. Thresholds with larger or degenerate code sizes but lower values than the ones shown in Fig.~\ref{fig:threshold} are neglected.

\begin{table*}[htbp!]
\centering
\begin{tabular}{|p{0.1\linewidth}|p{0.1\linewidth}||p{0.1\linewidth}|p{0.125\linewidth}|}
 \hline
\multicolumn{4}{|c|}{cyclic architecture thresholds} \\
 \hline
\multicolumn{2}{|c||}{active} & \multicolumn{2}{c|}{static} \\
 \hline
 $(n,m,j)$ & $\eta_c$ & $(n,m)$ & $\eta_c$ \\
 \hline
 (2,2,1)    & 0.0261    & (3,1)   & 0.0076 \\
 (2,3,1)    & 0.0495    & (3,2)   & 0.0381 \\
 (2,4,2)    & 0.0572    & (4,2)   & 0.0546 \\
 (3,3,1)    & 0.0753    & (4,3)   & 0.0675 \\
 (4,3,1)    & 0.083     & (5,3)   & 0.0856 \\
 (4,4,2)    & 0.097     & (6,3)   & 0.0962 \\
 (5,4,1)    & 0.1087    & (7,3)   & 0.1016 \\
 (6,4,1)    & 0.1172    & (7,4)   & 0.105  \\
 (7,4,1)    & 0.1217    & (8,4)   & 0.1143 \\
   -        & -         & (9,4)   & 0.1216 \\
 & & & \\
 & & & \\
 & & & \\
 \hline
\end{tabular}
\hspace{0.0001cm}
\begin{tabular}{|p{0.1\linewidth}|p{0.1\linewidth}||p{0.1\linewidth}|p{0.1\linewidth}|}
 \hline
\multicolumn{4}{|c|}{minimal architecture thresholds} \\
 \hline
\multicolumn{2}{|c||}{active} & \multicolumn{2}{c|}{static} \\
 \hline
 $(n,m,j)$ & $\eta_c$ & $(n,m)$ & $\eta_c$ \\
 \hline
 (2,3,2)    & 0.026     & (4,2)   & 0.02   \\
 (2,4,3)    & 0.0318    & (5,2)   & 0.038  \\
 (3,3,1)    & 0.044     & (5,3)   & 0.0425 \\
 (4,3,1)    & 0.0618    & (6,3)   & 0.0610 \\
 (4,4,2)    & 0.071     & (7,3)   & 0.074  \\
 (6,4,2)    & 0.089     & (8,3)   & 0.0818 \\
 (7,4,1)    & 0.098     & (11,3)  & 0.0925 \\
 (8,4,1)    & 0.1043    & (10,4)  & 0.096  \\
 (9,4,1)    & 0.1083    & (11,4)  & 0.1025 \\
 (10,4,1)   & 0.111     & (12,4)  & 0.1082 \\
(10,5,2)    & 0.1183    & (13,4)  & 0.113  \\
(11,5,2)    & 0.1225    & (14,4)  & 0.1168 \\
    -       &   -       & (15,4)  & 0.12   \\
 \hline
\end{tabular}
\vspace{0.1cm}
\caption{Threshold values $\eta_c$ plotted in Fig.~\ref{fig:threshold} and their corresponding parameter settings. $n,m$ are the QPC code parameters, and $j$ is the feed-forward parameter for the active encoded measurements.}
\label{tb:thresholds}
\end{table*}

\end{document}